\documentclass[letterpaper,12pt]{article}
    
\usepackage{blindtext} 

\usepackage{dirtytalk}	
\usepackage{lscape} 

	\usepackage{soul} 

    \usepackage{caption} 
    \usepackage[skip=0.5ex]{subcaption} 
    \usepackage{graphicx} 
    \usepackage{float} 
    
    \usepackage{printlen}
    \usepackage{amsmath}
    \numberwithin{figure}{section}

    \usepackage{makecell} 
    
    \usepackage{longtable} 
    \usepackage{tabularx} 
    \newcolumntype{L}[1]{>{\raggedright\arraybackslash}p{#1}}
    \newcolumntype{C}[1]{>{\centering\arraybackslash}p{#1}}
    \newcolumntype{R}[1]{>{\raggedleft\arraybackslash}p{#1}}
    \usepackage{relsize} 
    \usepackage[flushleft]{threeparttable}
    \usepackage{threeparttablex}
    \usepackage{multirow} 
    \usepackage{booktabs} 
    \usepackage{adjustbox} 
    \numberwithin{table}{section}

    \usepackage{abstract} 
    
    \usepackage{titlesec} 

    \usepackage{authblk} 

	\usepackage{datetime} 
    \newdateformat{usvardate}{
  	\monthname[\THEMONTH] \ordinal{DAY}, \THEYEAR}

  	\usepackage[bottom]{footmisc} 
    
 \usepackage{comment}
    
    \usepackage{fancyhdr}
    \fancyheadoffset{0cm}
    \providecommand{\keywords}[1]{\textbf{\textit{Keywords---}} #1}

    \usepackage{amssymb}
    \usepackage{amsthm}
    \usepackage{newtxmath}
    \numberwithin{equation}{section}
    \makeatletter
    \@addtoreset{equation}{section}
    \makeatother

    \usepackage{microtype} 
    \usepackage[utf8]{inputenc}
    \usepackage{newtxtext} 
  
	\usepackage{setspace} 
    \usepackage{lineno,xcolor} 	

	\usepackage[top=1.5in, bottom=1.5in, left=1in, right=1in]{geometry}

    \usepackage{graphicx} 
    \usepackage{float} 

	\usepackage{hyperref} 

    \usepackage{chicago} 

\begin{document}



\thispagestyle{empty} 




  \title{\vspace{-15mm}\fontsize{21pt}{10pt}\selectfont\textbf{Investigating Mode Effects in Interviewer Variances Using Two Representative Multi-mode Surveys}\footnote{Wenshan Yu, Survey Research Center at the Institute for Social Research, The University of Michigan; Dr. Michael R. Elliott, Survey Research Center at the Institute for Social Research and Biostatistics Department, The University of Michigan; Dr. Trivellore E. Raghunathan, Survey Research Center at the Institute for Social Research and Biostatistics Department, The University of Michigan. A significant portion of this paper will appear in a forthcoming issue of Survey Methodology.}}

	\date{Submitted: \usvardate\today}

\makeatletter
\renewcommand{\@maketitle}{
\newpage
 \null
 \vskip 2em%
 \begin{center}%
  {\LARGE \@title \par}%
 \end{center}%
 \par} \makeatother

\maketitle 






\begin{abstract}
\noindent 
As mixed-mode designs become increasingly popular, their effects on data quality have attracted much scholarly attention. Most studies focused on the bias properties of mixed-mode designs; few of them have investigated whether mixed-mode designs have heterogeneous variance structures across modes. While many characteristics of mixed-mode designs, such as varied interviewer usage, systematic differences in respondents, varying levels of social desirability bias, among others, may lead to heterogeneous variances in mode-specific point estimates of population means, this study specifically investigates whether interviewer variances remain consistent across different modes in mixed-mode studies. To address this research question, we utilize data collected from two distinct study designs. In the first design, when interviewers are responsible for either face-to-face or telephone mode, we examine whether there are mode differences in interviewer variances for 1) sensitive political questions, 2) international items, 3) and item missing indicators on international items, using the Arab Barometer wave 6 Jordan data. In the second design, we draw on Health and Retirement Study (HRS) 2016 core survey data to examine the question on three topics when interviewers are responsible for both modes. The topics cover 1) the CESD depression scale, 2) interviewer observations, and 3) the physical activity scale. To account for the lack of interpenetrated designs in both data sources, we include respondent-level covariates in our models. We find significant differences in interviewer variances on one item (twelve items in total) in the Arab Barometer study; whereas for HRS, the results are three out of eighteen. Overall, we find the magnitude of the interviewer variances larger in FTF than TEL on sensitive items. We conduct simulations to understand the power to detect mode effects in the typically modest interviewer sample sizes. 
\end{abstract}

\setlength\parindent{.45in} \keywords{Interviewer Effects, Mode Effects, Mixed-mode Design, Multimode Study}

\doublespacing


\section{Introduction}

Interviewers play a central role in survey data collection. Depending on the mode and sampling design of data collection, they may need to list addresses to generate sampling frames, recruit respondents, ask survey questions, and record participants' responses. Therefore, from a total survey error framework, interviewers can affect survey data quality by generating or reducing coverage error, nonresponse error, measurement error, and processing error \cite{west2017explaining}. Most research examining interviewers' effects focuses on measurement error \cite{schuman1971effects,hanson1958influence,ehrlich1961age}, which can be further decomposed into a systematic part, the bias due to interviewers (when respondents alter answers either because of the presence of interviewers or their observable traits), and a random component, interviewer variance. This interviewer variance inflates the uncertainty of the estimates, sometimes to an even greater degree than the correlation induced by geographical clustering \cite{schnell2003separating}. This study focuses on determining the effect of different modes of data collection -- specifically telephone (TEL) versus face-to-face (FTF) -- on interviewer variances in mixed-mode surveys.

Interviewer variances were first studied in the context of face-to-face interviews \cite{kish1962studies}. When telephone surveys became an alternative to FTF interviews, researchers evaluated interviewer variances in telephone surveys and generally found that they were less substantial than those in personal surveys \cite{groves1986measuring,tucker1983interviewer,groves1979surveys}. Specifically, the intraclass correlation $\rho_{int}$, a common measure used to assess interviewer effects and defined by the ratio of interviewer variances to the total variance, ranged from 0.005 to 0.102 in FTF surveys, whereas those computed in centralized TEL surveys ranged from 0.0018 to 0.0184 \cite{groves1986measuring,groves1979surveys}. The finding is aligned with theoretical expectations, as interviewers in the centralized TEL setting are more closely monitored and supervised than field interviewers are \cite{schaeffer2010interviewers}. Since then, the research domain has received little scholarly attention. However, as mixed-mode designs become increasingly used, the subject of study calls for more research. There is a lack of first-hand evidence as the prior findings are mostly based on different surveys that employ one mode (FTF or TEL). Besides, mixed-mode surveys naturally provide an opportunity where the survey context and the questionnaires are highly comparable (if not the same) when comparing interviewer variances in both modes. Furthermore, depending on whether interviewers are responsible for both modes in mixed-mode surveys, interviewers can potentially carry their influence from one mode to another. These factors can lead to different results in comparing interviewer variances between modes. 

Investigating mode effects in interviewer variances is also useful to facilitate mixed-mode designs and serve as an indicator of data quality. First, quantifying mode-specific interviewer variance can help researchers to determine and choose the mode with low interviewer variance in a multimode design. The current state-of-the-art mixed-mode inference strategy focuses on the bias property of modes \cite{elliott2009effects,kolenikov2014evaluating}, but little was done to incorporate the potential heterogeneous variance structure \cite{suzer2013investigating}. Part of the reason is that little literature sheds light on the variance properties of mixed-mode designs \cite{vannieuwenhuyze2015mode}, especially what goes into the variances. Second, identifying the questions associated with large interviewer variance mode effects can inform how interviewer variance is generated and thus might be reduced. For example, researchers show that attitudinal, sensitive, ambiguous, complex, and open-ended questions are generally more vulnerable to interviewer effects \cite{schaeffer2010interviewers}, as those questions introduce more opportunities for the interviewer to help the respondents \cite{west2017explaining}. If sensitive questions only present a large interviewer effect in FTF but not in TEL, that may suggest the questions bring a burden to field interviewers. To address that, survey organizations can provide additional training to standardize how to ask the question or use other approaches [such as audio computer-assisted self-interviewing [ACASI] or the item count technique \cite{holbrook2010social}] to collect information for sensitive items. Third, in mixed-mode designs where interviewers are responsible for both modes, we can potentially find specific interviewers that have a large effect on responses in both modes or only in one mode, which provide the basis for real-time intervention and interviewer training at a more granular level.

In this paper, we consider two representative multi-mode studies: 1) the Arab Barometer Study (ABS) Wave 6 Jordan experiment and 2) the Health and Retirement Study (HRS) 2016. Drawing on both data sources, we consider mode effects in interviewer variances for interviewers in different countries, for different target populations, and for a variety of outcome variables. Additionally, the use of the two studies offers distinct perspectives for examining our research question. The ABS interviewer design is commonly used in surveys where different modes are managed by separate data collection agencies, resulting in different interviewers across modes. On the other hand, the HRS interviewer design, where the same interviewers are utilized in both modes, facilitates a more precise estimation of the differences in interviewer variances solely due to modes, by eliminating the portion of interviewer variances that result from using different interviewers across modes.

The remainder of this paper is organized as follows. In Section 2, we describe the study design and analytical strategy, and present the results using our first data source -- ABS. Section 3 introduces the second data source -- HRS, along with the corresponding analytical approach and the results pertaining to interviewer variance associated with the HRS data. In Section 4, we conduct a simulation study to illustrate the power to detect mode effects in interviewer variances using both the ABS and the HRS setup. Finally, in Section 5, we discuss the implications of our study.

\section{The Arab Barometer Study}
\subsection{Study Description}
The ABS is the largest repository of public opinion data in the Middle East and North Africa (MENA) region. In wave 6, it embedded a mode experiment in Jordan between March and April 2021, where participants were randomly assigned to either a personal interview or a TEL recontact interview. Center for Strategic Studies in Jordan conducted the field work using the 2015 Population and Housing Census as the sampling frame. They implemented an area probability sample stratified on governorate and urban-rural cleavages. Separate interviewers were used in the FTF and TEL interviews. The TEL-assigned households were initially recruited via FTF for a short 5-minute survey, and the majority of the survey items were asked approximately a week later in a telephone follow-up. In the FTF mode, 31 interviewers collected data from 1,193 respondents, while 13 interviewers interviewed 1,212 participants via phone. 

We focus on three types of outcome variables ($Y$): 1) sensitive political questions (6 items), 2) less sensitive international questions (3 items), and 3) whether reported do not know or refused to answer international relationship questions (3 items). Except for the item missing indicators, the other outcome variables were initially measured by four ordinal categories; we collapsed them into binary outcomes by setting the cutoff point in the middle (see the original and the collapsed categories in the Appendix A).

Outcome variables ($Y$) can be subject to two types of mode effects: 1) mode effects that lead to a shift in the means of outcome variables (referred to as mode effects in means) and 2) mode effects in interviewer variances. We consider, in total, $q$ interviewers collect information in only one of two modes (FTF and TEL) from $n$ sample units from a finite population. Interviewers also collect respondent-level covariates ($X$) that are predictive of the outcome variables ($Y$). The covariates ($X$) are assumed to be independent of any mode effects. We consider covariates ($X$) including respondents’ age, gender, marital status, household size, and regions in this paper.

\subsection{Analytical Strategy}
First, to illustrate the descriptive statistics of interviewer variation in the collected responses, we compute the between-interviewer standard deviation (SD) and the average within-interviewer SD. Specifically, we calculate the average proportions for each variable and interviewer ($\bar{y}_{(m)j}$). In the ABS setup, where interviewers are nested within each mode, these statistics are inherently mode-specific; therefore, we enclose $m$ in parentheses to emphasize this point. We then calculate the SD of these average proportions across interviewers, termed the between-interviewer SD. The within-interviewer SD ($v_j^{m}$) is derived from the responses collected by each interviewer. The average within-interviewer SD is computed as the mean of the within-interviewer SDs across all interviewers for each mode. We show the formula to compute the relevant statistics in \ref{descriptiveStat}, where $i$ indexes respondents, $j$ indexes interviewers, $m$ indexes modes, $n_{(m)j}$ reflects the number of interviews conducted by interviewer $j$ using mode $m$, $n_{m}$ represents the number of respondents in mode $m$, $n_j^m$ indicates the number of interviewers using mode $m$, and $y_{i(m)j}$ indicates the responses provided by respondent $i$ interviewed by interviewer $j$ using mode $m$. From the perspective of survey data collection agencies, a small SD between interviewers and a large average within-interviewer SD are desirable, as this may indicate an interviewer assignment that is close to random and minimal effects from interviewers on the collected responses. We report the statistics for both the covariates and the outcomes of interest. The statistics for the covariates can suggest interviewer selection effects, thereby highlighting the importance of considering the covariates in the final analytical model. The statistics for the outcome variables may provide initial evidence of the presence of interviewer effects and justify further investigation.

\begin{equation}
\label{descriptiveStat}
\begin{split}
 &\text{Average proportion per interviewer } \bar{y}_{(m)j} = \frac{\sum_i^{n_(m)j} y_{i(m)j}}{n_{(m)j}}\\
 &\text{Average proportion per mode } \bar{y}_m = \frac{\sum_i^{n_m} y_{i(m)j}}{n_m}\\
 &\text{Between-interviewer SD } = \sqrt{\frac{\sum_j^{n^m_{j}}(\bar{y}_{(m)j}-\bar{y}_m)^2}{n^m_j}}\\
 &\text{Within-interviewer SD } v^m_j = \sqrt{\frac{\sum_i^{n_{(m)j}} (\bar{y}_{i(m)j}-\bar{y}_{(m)j})^2}{n_{(m)j}}}\\
 &\text{Average within-interviewer SD } = \frac{\sum_j^{n^m_j} v^m_j}{n^m_j}
\end{split}
\end{equation}

To test whether interviewer variances are equal across modes, since all the outcome variables are binary, we fit the following probit model to each of the variables, where $m$ indexes modes ($f$ for FTF and $t$ for TEL), $M$ and $J_{j,j=1,...,q-1}$ are dummy variables (length of $n$) to indicate modes ($M=1$ for the FTF mode and $M=0$ for the TEL mode) and interviewers:

    \begin{equation}
    \label{Model:Arab_base}
    \begin{split}
         & Y^*_{ij(m)}=\beta_0 +\beta_1M_{i}+b_{j(m)} + \epsilon_{ij(m)}, \\
         & Y_{ij(m)}=1 \text{ if } Y^*_{ij(m)}>0 \text{ and } Y_{ij(m)}=0 \text{ if } Y^*_{ij(m)}\leq 0, \\
         & b_{j(m)} \sim N(0,\sigma^2_{m}),\\
         &\epsilon_{ij(m)} \sim N(0,1),\\
         &\sigma_f, \sigma_t \sim half-T(3,1) \text{ (for Bayesian modeling)},\\
         &\boldsymbol{\gamma}, \beta_0, \beta_1 \sim N(0, 10^6) \text{ (for Bayesian modeling)}
    \end{split}
    \end{equation}
      
In Model \ref{Model:Arab_base}, the interviewer random effects are represented as $b_{j(m)}$ as interviewers are nested within the modes. Our research question, \say{Are interviewer variances equal between modes in a randomized mixed-mode design?} is addressed by evaluating if $\alpha = log(\sigma_f) - log(\sigma_t)$ is equal to zero for each variable in Model \ref{Model:Arab_base}. To determine this, we examine if the 95\% confidence or HPD credible intervals of $\alpha$ include zero. If the intervals do not include zero for some variables, it suggests that the interviewer variances are not equal between modes for those variables.

By fitting \ref{Model:Arab_base}, we can also obtain estimates of mode effects ($\beta_1$) for each variable by computing  and testing if the quantity differs from 0. Note that the estimates may include some mode selection effects; despite the random mode assignment, differential nonresponse can happen across the modes \cite{west2013interviewer}.

Suppose evidence suggests that $\alpha \neq 0$, we then consider whether the mode-specific interviewer variance is spurious due to the lack of interpenetrated designs by adding respondent-level covariates ($x_{si}$, where $s$ denotes covariate $s$) to Model \ref{Model:Arab_base}:

    \begin{equation}
    \label{Model:Arab_addX}
    \begin{split}
         & Y^*_{ij(m)}=\beta_0 +\beta_1M_{i}+b_{j(m)}+ \sum^S_s \gamma_sx_{si}+\epsilon_{ij(m)}, \\
         & Y_{ij(m)}=1 \text{ if } Y^*_{ij(m)}>0 \text{ and } Y_{ij(m)}=0 \text{ if } Y^*_{ij(m)}\leq 0, \\
         & b_{j(m)} \sim N(0,\sigma^2_{m}),\\
         &\epsilon_{ij(m)} \sim N(0,1),\\
         &\sigma_f, \sigma_t \sim half-T(3,1) \text{ (for Bayesian modeling)},\\
         &\boldsymbol{\gamma}, \beta_0, \beta_1 \sim N(0, 10^6) \text{ (for Bayesian modeling)}
    \end{split}
    \end{equation}

We implement the models using both likelihood (Proc Nlmixed) and Bayesian approaches (Proc MCMC) in the SAS programming language. In the likelihood approach, we take log transformation on $\sigma^2_f$ and $\sigma^2_t$ to stabilize the variance of the parameters and improve the coverage property. We compute the variance of the estimated $\alpha$ using the delta method, given by $var(\alpha) = \frac{1}{4} var(log(\sigma^2_f))+\frac{1}{4}var(log(\sigma^2_t))$ (see the derivations in the Appendix B), then use a normal distribution to estimate the 95\% confidence interval. In the Bayesian approach, we use one chain with 200,000-300,000 draws, depending on the autocorrelation and effective sample size, and select every 100th value as the thinning rate. For the ease of illustration, we only report the results of the model with covariates added and estimated using Bayesian modeling (Model \ref{Model:Arab_addX}) in the later section. 

\subsection{Results}
\subsubsection{Descriptive Statistics}
\begin{figure}
\centering
    \includegraphics[width=\textwidth]{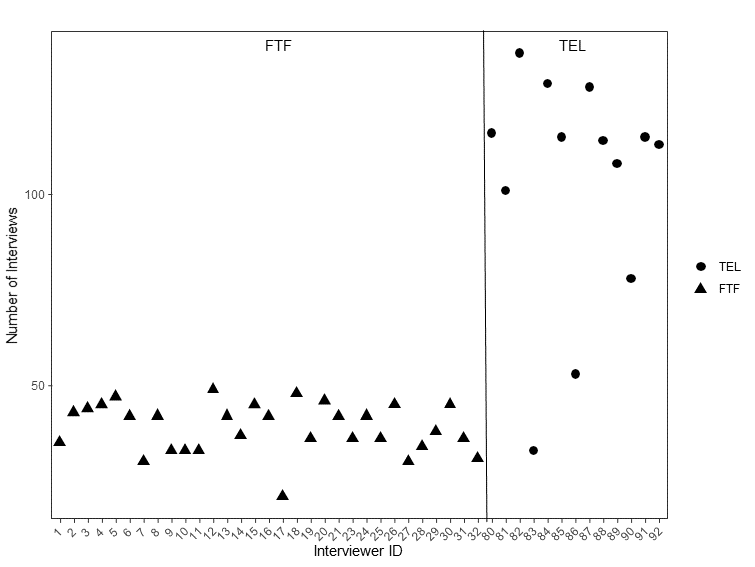}
    \caption{Interviewer Workloads Per Mode in the Arab Barometer Study}
    \label{fig:ArabIntWorkload}
\end{figure}

We assume interviewers are interchangeable in this paper. To partly evaluate this assumption, we present the interviewer workloads in the FTF and TEL modes in the ABS in Figure \ref{fig:ArabIntWorkload}. In Figure \ref{fig:ArabIntWorkload}, we note that in the FTF mode, each interviewer conducts a similar number of interviews. In contrast, both the mean and the variation in the number of interviews per interviewer are larger and more variable in the TEL mode.

We report mode-specific sample means, between-interviewer standard deviations (SDs), and average within-interviewer SDs in Table \ref{tab:descriptiveABS}. From Table \ref{tab:descriptiveABS}. First, we observe that for sensitive political questions, the average proportions reported via telephone (TEL) are generally higher than those reported in face-to-face interviews (FTF), suggesting that TEL may be associated with more positive reporting. Second, between-interviewer SDs in FTF are generally larger than those in TEL for most outcomes, while the average within-interviewer SD is larger in TEL than in FTF for sensitive political questions and missing indicators. This provides some initial evidence that interviewers seem to have a larger effect in FTF than in TEL. We provide the distribution of the outcome variables per interviewer in Appendix C.

\begin{landscape}
\begingroup\fontsize{11pt}{11pt}\selectfont
\begin{longtable}{lrrrrrr}
\caption{Distribution of Outcome variables in the Arab Barometer Study across Interviewers by Modes} \\ 
  \hline
  \hline
Questions & Mean (FTF) & Mean (TEL) & \makecell{Between\\ interviewer\\ SD (FTF)} & \makecell{Between\\ interviewer\\ SD (TEL)} & \makecell{Average\\ Within\\ interviewer\\ SD (FTF)} & \makecell{Average\\ Within\\ interviewer\\ SD (TEL)} \\ 
  \hline
\multicolumn{7}{c}{Sensitive political questions}\\
\hline
 \makecell{1. Freedom of the media} & 0.403 & 0.588 & 0.191 & 0.117 & 0.455 & 0.480 \\ 
 \makecell{2. trust in government} & 0.356 & 0.533 & 0.165 & 0.122 & 0.455 & 0.487 \\ 
 \makecell{3. trust in courts} & 0.594 & 0.770 & 0.139 & 0.123 & 0.477 & 0.398 \\ 
 \makecell{4. satisfied with healthcare} & 0.491 & 0.592 & 0.155 & 0.071 & 0.482 & 0.489 \\ 
 \makecell{5. performance on inflation} & 0.140 & 0.243 & 0.146 & 0.142 & 0.291 & 0.406 \\ 
  \makecell{6. performance during COVID-19} & 0.402 & 0.576 & 0.171 & 0.161 & 0.464 & 0.470 \\ 
  \hline
\multicolumn{7}{c}{International Questions}\\
\hline
\makecell{7. favorable of the United States} &  0.394 & 0.415 & 0.187 & 0.189 & 0.467 & 0.459 \\ 
 \makecell{8. favorable of Germany} & 0.488 & 0.560 & 0.224 & 0.186 & 0.464 & 0.464 \\ 
 \makecell{9. favorable of China} & 0.468 & 0.507 & 0.207 & 0.203 & 0.470 & 0.463 \\ 
 \hline
\multicolumn{7}{c}{Whether missing on international questions (constructed)}\\
\hline
 \makecell{10. missing on favorable of the\\ United States}& 0.253 & 0.297 & 0.235 & 0.158 & 0.341 & 0.425 \\ 
 \makecell{11. missing on favorable of Germany} & 0.320 & 0.381 & 0.247 & 0.199 & 0.384 & 0.442 \\ 
 \makecell{12. missing on favorable of China} & 0.283 & 0.329 & 0.252 & 0.180 & 0.359 & 0.431 \\ 
   \hline
\hline
\label{tab:descriptiveABS}
\end{longtable}
\endgroup
\end{landscape}

We show unweighted sample characteristics in the FTF and the TEL modes in Table \ref{tab:ArabRespDistPerMode}. Under the randomized mixed-mode design, the Jordan sample is roughly balanced on key demographic and socioeconomic variables (age, gender, education, marital status, household size, and region) across modes. However, there are slightly more males (0.55 vs 0.50) respondents in the TEL mode relative to the FTF mode, possibly due to differential nonresponse. We note that for these covariates, the between-interviewer SD in FTF is usually much larger than that in TEL, suggesting potentially larger selection effects in FTF, since we assume the covariates are not susceptible to measurement error.

\begin{landscape}
\begin{table}[ht]
\centering
\begin{threeparttable}
\caption{Distribution of Sample Characteristics of the Arab Barometer Study across Interviewers by Modes} 
\label{tab:ArabRespDistPerMode}
\begin{tabular}{lcccccc}
  \hline
Respondent Variables & Mean (FTF)  &
Mean (TEL) & \makecell{Between\\ interviewer\\ SD (FTF)} & \makecell{Between\\ interviewer\\ SD (TEL)} & \makecell{Average\\ Within\\ interviewer\\ SD (FTF)} & \makecell{Average\\ Within\\ interviewer\\ SD (TEL)} \\ 
  \hline
   Age 18-24  &0.166 & 0.164 & 0.085 & 0.039 & 0.361 & 0.369 \\ 
   Age 25-34  &0.226 & 0.203 & 0.072 & 0.038 & 0.415 & 0.402 \\
   Age 35-44  & 0.227 & 0.215 & 0.088 & 0.052 & 0.412 & 0.408 \\
   Age 45-54  &0.199 & 0.219 & 0.069 & 0.031 & 0.394 & 0.414 \\ 
   Age 55+  &0.183 & 0.198 & 0.071 & 0.032 & 0.381 & 0.399 \\ 
   Male &0.497 & 0.549 & 0.291 & 0.041 & 0.369 & 0.499 \\
   Less than secondary education &0.345 & 0.337 & 0.125 & 0.106 & 0.463 & 0.463 \\
   Secondary education &0.365 & 0.357 & 0.098 & 0.082 & 0.477 & 0.474 \\ 
   Higher than secondary education &0.290 & 0.307 & 0.101 & 0.051 & 0.445 & 0.461 \\ 
   Unmarried &0.238 & 0.264 & 0.106 & 0.063 & 0.412 & 0.438 \\ 
   Married &0.693 & 0.684 & 0.082 & 0.062 & 0.459 & 0.463 \\ 
   Divorced, widows, separated &0.069 & 0.053 & 0.044 & 0.024 & 0.230 & 0.219 \\
   Household size: Less than 3 &0.208 & 0.222 & 0.083 & 0.040 & 0.399 & 0.416 \\
   Household size: 4-5 &0.345 & 0.349 & 0.081 & 0.061 & 0.475 & 0.475 \\ 
   Household size: 6-7 &0.281 & 0.288 & 0.079 & 0.072 & 0.447 & 0.449 \\ 
   Household size: 8+ &0.165 & 0.141 & 0.089 & 0.056 & 0.353 & 0.330 \\
   Region: Central &0.523 & 0.509 & 0.154 & 0.255 & 0.482 & 0.429 \\ 
   Region: North &0.261 & 0.282 & 0.101 & 0.188 & 0.424 & 0.388 \\
   Region: South &0.216 & 0.209 & 0.175 & 0.119 & 0.333 & 0.367 \\
   \hline
\end{tabular}
\end{threeparttable}
\end{table} 
\end{landscape}

\subsubsection{Mode effects in Means and Interviewer Variances}
This section reports the modeling results that incorporate respondent information (Model \ref{Model:Arab_addX}) using Bayesian estimation in Table \ref{tab3:ArabBarometerResults_addCov}. With respect to the mode effects in means, we observe negative estimates for all sensitive items. For example, the probability of an unmarried male participant aged 18-24, with higher than secondary education, living in a household with fewer than three individuals, and residing in the North region of Jordan, reporting that media freedom is guaranteed to a great or medium extent, decreases by 17.9\% when interviewed via face-to-face (FTF) methods compared to telephone (TEL) interviews. The 17.9\% is calculated using $\phi(\beta_0+\beta_1+\sum_s^S \gamma_sx_{si})\beta_1$, where $\phi$ is the pdf of a standard normal distribution and S is the number of covariates ($x$). The estimates of $\gamma_s$ are not provided in the paper but can be provided upon request. The negative mode effects in means suggest that respondents expressed lower opinions of the government when answering FTF interviews, which could be more honest responses given Jordan's authoritarian regime. Table \ref{tab3:ArabBarometerResults_addCov} also indicates that missing rates for international questions are lower in FTF interviews compared to TEL interviews (though this is not statistically significant at the 0.05 level). We did not incorporate sample weights in the analysis as our focus of inference is repeated sampling under the same survey design.

Next, we turn our attention to the interviewer variances. Firstly, the magnitude of interviewer variances is generally large in the ABS. For sensitive political questions, the interviewer variances range from 0.018 to 0.393 (Table \ref{tab3:ArabBarometerResults_addCov}). Previous literature examining interviewer effects usually reported interviewer intraclass correlation ($\rho_{int}$) to reflect the proportion of variance due to interviewers. To compute mode-specific $\rho_{m,int}$, we can use the formula $\rho_{m,int} = \frac{var_{m,int}}{1+var_{m,int}}$, since the residual variance in the probit model is 1. Consequently, the previously mentioned results correspond to $\rho_{int}$ ranging from 0.018 to 0.282. As a reference, based on the literature, a value of $\rho_{int}$ below 0.01 is considered small, while a value higher than 0.12 is regarded as large \cite{west2010much}. In Table \ref{tab3:ArabBarometerResults_addCov}, we observe that $\rho_{f,int}$ and $\rho_{t,int}$ can vary substantially for the same outcome. For example, for satisfaction with healthcare, $\rho_{f,int}$ is 0.125, while $\rho_{t,int}$ is 0.029. It is important to consider these differences when using the $\rho_{m,int}$ values to calculate the effective sample sizes associated with a specific data collection mode.

For one sensitive item, performance in the healthcare system, we observe marginally significant difference in interviewer variances in Table \ref{tab3:ArabBarometerResults_addCov} using Bayesian estimation (and significant using likelihood estimation, see Appendix D). In this item, the estimates of interviewer variances are considerably larger in the FTF mode. For 5 out of 6 sensitive items, FTF interviewer variances are somewhat larger than TEL interviewer variances. The differences are not statistically significant, possibly due to the limited power determined by the small number of interviewers in this study. The larger interviewer variances in FTF are consistent with theoretical expectations, as interviewers may exhibit greater heterogeneity in administering sensitive questions and establishing rapport with respondents during in-person interviews. 

Counterintuitively, for substantive responses to nonsensitive international attitude questions (items 7-9), the interviewer variance estimates are generally larger in TEL compared to FTF (not significantly). The interviewer variances of whether reporting \say{don't know} or refusing to answer the nonsensitive international questions are larger in FTF than in TEL (significant on the first item). This finding may be because interviewers assigned to FTF mode tried to persuade respondents to give substantive answers, and whether the persuasion happens or is successful can differ by interviewers.

\begin{landscape}
\begin{ThreePartTable}
\begin{TableNotes}
\small
    \item Notes: Significant results are marked in bold. $\beta_1$ refers to the mode effect estimates in means. $\alpha_1$ refers to the mode effect estimates in interviewer variances. $\sigma^2_f$ is the FTF interviewer variances. $\sigma^2_t$ is the TEL interviewer variance. $\rho_{f,int}$ and $\rho_{t,int}$ are interviewer intraclass correlation in FTF and TEL, respectively.   
\end{TableNotes}
\begingroup\fontsize{11pt}{11pt}\selectfont
\begin{longtable}{
@{}
>{\raggedright}p{0.30\linewidth}
>{\raggedright}p{0.09\linewidth}
>{\raggedright}p{0.09\linewidth}
>{\raggedright}p{0.09\linewidth}
>{\raggedright}p{0.09\linewidth}
>{\raggedright}p{0.09\linewidth}
p{0.09\linewidth}
@{}}
\caption{Interviewer Variances Per Mode for Selected Items in the Arab Barometer Study Adjusting for Covariates Using Bayesian Estimation} \\ 
  \hline
Questions & $\sigma^2_f$ & $\sigma^2_t$ & $\rho_{f,int}$ & $\rho_{t,int}$ &$\alpha$ & $\beta_1$ \\
\hline
\hline
\multicolumn{7}{c}{Sensitive political questions}\\
\hline
\makecell{1. Freedom of the media} & 0.252 [0.122, 0.428] & 0.135 [0.036, 0.284] & 0.201 [0.109, \:0.3] & 0.119 [0.035, 0.221] & 0.355 [-0.223, 0.898] & \textbf{-0.526 [-0.795, -0.222]} \\
\hline
\makecell{2. trust in government} & 0.188 [0.083, 0.322] & 0.127 [0.029, 0.275] & 0.158 [0.077, 0.244] & 0.113 [0.028, 0.216] & 0.239 [-0.382, 0.838] & \textbf{-0.504 [-0.768, -0.238]} \\  
\hline
\makecell{3. trust in courts} & 0.113 [0.038, 0.201] & 0.214 \: [0.05, 0.445] & 0.102 [0.037, 0.167] & 0.176 [0.048, 0.308] & -0.291 [-0.94, 0.318] & \textbf{-0.555 [-0.881, -0.273]} \\
\hline
\makecell{4. satisfied with healthcare} & 0.143 [0.051, 0.251] & 0.03 \qquad [0,\qquad 0.075] & 0.125 [0.049, 0.201] & 0.029 \qquad [0, \qquad 0.07] & 0.906 [-0.054, 1.758] & \textbf{-0.278 [-0.475, -0.085]} \\ 
\hline
\makecell{5. performance on inflation} & 0.393 [0.153, 0.672] & 0.204 [0.051, 0.435] &0.282 [0.133, 0.402] & 0.169 [0.049, 0.303] & 0.361 [-0.275, 0.927] & \textbf{-0.523[-0.861, -0.153]} \\
\hline
\makecell{6. performance during COVID-19} & 0.202 [0.084, 0.34] & 0.224\qquad [0.07, 0.443] &0.168 [0.077, 0.254] & 0.183 [0.065, 0.307] & -0.026 [-0.602, 0.508] & \textbf{-0.51 \hspace{1cm} [-0.841, -0.205]} \\
\hline
\multicolumn{7}{c}{International Questions}\\
\hline
\makecell{7. favorable of the United States} & 0.198 [0.074, 0.34] & 0.362  [0.104, 0.719] & 0.165 [0.069, 0.254] & 0.266 [0.094, 0.418] & -0.278 [-0.841, 0.282] & -0.057 [-0.45, 0.318] \\ 
\hline
\makecell{8. favorable of Germany} & 0.292 \hspace{1cm}[0.12, 0.514] & 0.33 \hspace{1cm}[0.092, 0.663] & 0.226 [0.107, 0.339] & 0.248 [0.084, 0.399] & -0.037 [-0.603, 0.548] & -0.147 [-0.551, 0.236] \\ 
\hline
\makecell{9. favorable of China} & 0.205 [0.083, 0.361] & 0.378 [0.116, 0.787]  & 0.17 \hspace{1cm} [0.077, 0.265] & 0.274 [0.104, 0.44] & -0.282 [-0.869, 0.245] & -0.15 [-0.549, 0.19] \\
\hline
\multicolumn{7}{c}{Whether missing on international questions (constructed)}\\
\hline
\makecell{10. missing on favorable of the\\ United States} & 0.995 \hspace{1cm}[0.48, 1.71] & 0.343 [0.104, 0.668] & 0.499 [0.324, 0.631] & 0.255 [0.094, \hspace{1cm}0.4] & \textbf{0.557 [0.014, 1.121]} & -0.298 [-0.805, 0.172] \\ 
\hline
\makecell{11. missing on favorable of Germany} & 0.844 [0.404, 1.324] & 0.464 \hspace{1cm}[0.16, 0.857] & 0.458 [0.288, 0.57] & 0.317 [0.138, 0.461] & 0.324 [-0.169, 0.839] & -0.287 (0.24) [-0.765, 0.149] \\ 
\hline
\makecell{12. missing on favorable of China} & 0.936 [0.434, 1.552] & 0.452  [0.118, 0.933] & 0.483 [0.303, 0.608] & 0.311 [0.106, 0.483] & 0.398 [-0.134, 0.949] & -0.244 [-0.73, 0.229] \\
\hline
\hline
\insertTableNotes
\label{tab3:ArabBarometerResults_addCov}
\end{longtable}
\endgroup
\end{ThreePartTable}
\end{landscape}

\section{Health and Retirement Study 2016}
\subsection{Study Description}
The HRS is a longitudinal panel study that surveys people over age 50 (and their spouses) in the United States. It is conducted biennially, started in 1992, and has studied more than 43,000 people \cite{fisher2018overview}. The HRS is sponsored by the National Institute on Aging (grant number NIA U01AG009740) and is conducted by the University of Michigan. The HRS sample was drawn using a multistage, national area-clustered probability sample frame \cite{heeringa1995technical}. Since 2006, The HRS has initiated the rotation of enhanced FTF and TEL across waves for participants, unless they are older than 80 years, newly recruited into the sample, or spouses of another HRS participant, in which case they rotate between regular FTF and enhanced FTF. In this study, we are interested in analyzing the HRS 2016 data, when the Late Baby Boomers (LBB) cohort was added to replenish the HRS sample. Although not every interviewer collects data in both modes, under the HRS design, interviewers are responsible for data collection in both FTF and TEL modes. The HRS 2016 was fielded from April 2016 to April 2018, with a sample size of 20,912 [response rate: 82.8\%, \cite{HRS2023}]. In our analytical sample, we excluded respondents who were missing data on mode indicators, interviewer IDs, and covariates, resulting in a sample size of 20,868.  

We consider four types of outcome variables in the HRS study, including 1) nine items of the Center for Epidemiologic Studies Depression Scale (CESD), 2) six items of interviewer observations, and 3) a three-item physical activity scale (see the Appendix E for the question wordings, the original response categories and categories used in the study). We consider nine respondent-level covariates ($X$), including age, sex, race / ethnicity, interview language, education, whether respondents are coupled and working. All participants are included in our sample, unless they are missing data in either the outcome or predictor variables. Missing rates for predictor variables are minor, and those for outcome variables are less than 0.05.

\subsection{Analytical Strategy}
Similar to the descriptive statistics reported in the ABS, we report the between-interviewer SD and the average within-interviewer SD to gain an intuitive understanding of the interviewer effects in the outcome variables examined in the HRS.

Next, we fit multilevel models to each of the outcome variables using the same notation as in Model \ref{Model:Arab_base}. Unlike the ABS, interviewers are not nested in model hence a single interviewer can interview in both modes, and thus interviewer effects can be correlated across modes. Therefore we posit a bivariate normal model for the interviewer effects: 
 
    \begin{equation}
    \label{Model:hrs_binary}
    \begin{split}
         & Y^*_{ijm}=\beta_0 +\beta_1M_{i}+b_{jm}+ \sum^S_s \gamma_sx_{si}+\epsilon_{ijm}, \\
         & Y_{ijm}=1 \text{ if } Y^*_{ijm}>0 \text{ and } Y_{ijm}=0 \text{ if } Y^*_{ijm}\leq 0, \\
         & \begin{pmatrix}
         b_{jf} \\ b_{jt}
        \end{pmatrix} \sim N \Bigg( \begin{pmatrix}
        0 \\ 0
        \end{pmatrix}, \begin{pmatrix}
        \sigma^2_{f} & \rho \sigma_{f}  \sigma_{t} \\
        \rho \sigma_{f}\sigma_{t} & \sigma^2_{t} 
        \end{pmatrix} \Bigg),\\
         &\epsilon_{ijm} \sim N(0,1),\\
         &\sigma_f, \sigma_t \sim half-T(3,1) \text{ (for Bayesian modeling)},\\
         &\rho \sim U(-1,1)  \text{ (for Bayesian modeling)},\\
         &\boldsymbol{\gamma}, \beta_0, \beta_1 \sim N(0, 10^6) \text{ (for Bayesian modeling).}
    \end{split}
    \end{equation}
      
Similarly, we use $\alpha = log(\sigma_f) - log(\sigma_t)$ as a metric to answer our research question. To test if $\alpha$ is equal to zero for each variable, we assess if the 95\% credible intervals or confidence intervals include zero. Additionally, to control for interviewer selection effects, we include respondent-level covariates as fixed effects in the model.

We apply the Fisher Z transformation ($z = \frac{1}{2}ln(\frac{1+\rho}{1-\rho})$) when constructing the 95\% confidence interval for $\rho$ in the likelihood approach. We calculate the variance of $\alpha$ using the delta method, given by $var(\alpha) = \frac{1}{4} var(log(\sigma^2_f)) + \frac{1}{4} var(log(\sigma^2_t)) - \frac{1}{2} cov(log(\sigma^2_f), log(\sigma^2_t))$, which is slightly different from the ABS (see the derivations in Appendix B).

\subsection{Results}
\subsubsection{Descriptive Statistics}
\begin{figure}
\centering
    \includegraphics[width=\textwidth]{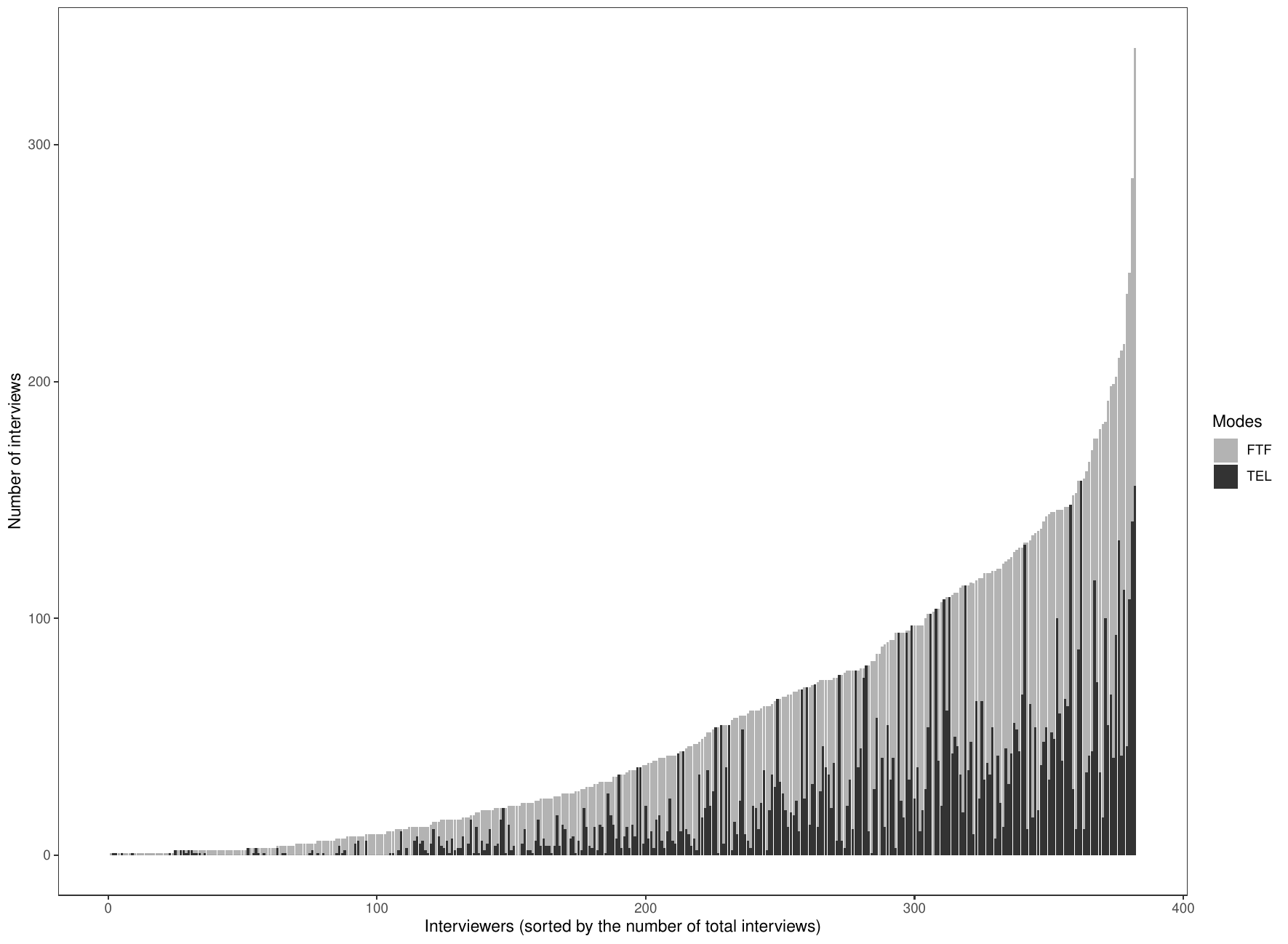}
    \caption{Interviewer Workloads Per Mode in the Health and Retirement Study}
    \label{fig:HRSIntWorkload}
\end{figure}

 First, we illustrate the interviewer load in Figure \ref{fig:HRSIntWorkload}. In HRS 2016, 382 interviewers were employed for data collection. The number of interviews conducted in FTF and TEL is very different across interviewers. Eighty-two (21.5\%) interviewers exclusively conducted telephone interviews, while thirty-seven (9.7\%) solely conducted in-person interviews. The remaining 263 (68.9\%) interviewers conducted both types of interviews. All interviews are included in the analysis, although estimation of the covariances between the FTF and TEL effects within interviewer are limited to the subsample of interviewers who conducted both types of interviews. 

Second, we present unweighted sample characteristics for both FTF and TEL modes in Table \ref{tab:HRSDistPerMode}. Compared to TEL respondents, a higher proportion of FTF respondents are under 60 or over 80 years old, belong to minority groups, are not in a relationship, have not completed high school, and are currently employed. This unbalanced sample distribution underscores the importance of including demographic and socioeconomic status variables in the analytical model when analyzing interviewer effects. Comparing the statistics from HRS to those from ABS, we note that in the HRS, the between-interviewer SDs are generally higher and the average within-interviewer SDs are generally lower. This suggests that interviewer selection effects are potentially a larger threat when analyzing interviewer variance in the HRS. This is consistent with our expectations, as randomized mode assignment is applied in ABS but not in HRS.   

\begin{landscape}
\begin{table}[h]
\centering
\begin{threeparttable}
\caption{Distribution of Sample Characteristics in the Health and Retirement Study across Interviewers by Modes} 
\label{tab:HRSDistPerMode}
\begin{tabular}{lcccccc}
  \hline
Respondent Characteristics & Mean (FTF)  &
Mean (TEL) & \makecell{Between-\\interviewer\\ SD (FTF)} & \makecell{Between-\\interviewer\\ SD (TEL)} & \makecell{Average\\ Within-\\interviewer \\SD (FTF)} & \makecell{Average\\ Within-\\interviewer \\SD (TEL)} \\ 
  \hline
Age: less than 60 & 0.449 & 0.305 & 0.359 & 0.315 & 0.294 & 0.363 \\ 
Age: 60-69 &  0.188 & 0.343 & 0.156 & 0.239 & 0.260 & 0.398 \\ 
Age: 70-79 & 0.181 & 0.287 & 0.150 & 0.255 & 0.240 & 0.342 \\ 
Age: 80+ & 0.182 & 0.066 & 0.185 & 0.151 & 0.232 & 0.163 \\
Currently Working &  0.368 & 0.329 & 0.280 & 0.265 & 0.426 & 0.427 \\ 
Male & 0.414 & 0.415 & 0.177 & 0.198 & 0.503 & 0.503 \\ 
Spanish-speaking Hispanic & 0.091 & 0.085 & 0.194 & 0.212 & 0.087 & 0.072 \\  
English-speaking Hispanic & 0.077 & 0.074 & 0.158 & 0.146 & 0.198 & 0.189 \\
Black & 0.219 & 0.200 & 0.274 & 0.246 & 0.315 & 0.344 \\ 
White &  0.613 & 0.641 & 0.315 & 0.300 & 0.376 & 0.397 \\
Coupled & 0.601 & 0.632 & 0.260 & 0.275 & 0.431 & 0.428 \\ 
Education:less than 12 years & 0.203 & 0.188 & 0.190 & 0.225 & 0.348 & 0.324 \\
Education:12 years &  0.303 & 0.290 & 0.188 & 0.218 & 0.420 & 0.416 \\  
Education:13-15 years &  0.259 & 0.268 & 0.196 & 0.200 & 0.406 & 0.421 \\
Education:16 years + & 0.249 & 0.262 & 0.214 & 0.220 & 0.391 & 0.374 \\  
   \hline
\end{tabular}
\end{threeparttable}
\end{table}    
\end{landscape}

Next, we present the descriptive statistics of the HRS, including mode-specific sample means, between-interviewer standard deviation (SD), and average within-interviewer SD in Table \ref{tab:descriptiveHRS}. First, for the CESD scale, the prevalence rates are generally higher in face-to-face (FTF) interviews than in telephone (TEL) interviews, suggesting that FTF may be associated with more honest reporting. Second, interviewers report FTF respondents as more attentive, understanding questions better, less cooperative, having less difficulty remembering but more difficulty hearing things, and interviewed with higher quality. Third, the magnitude of the between-interviewer SD appears larger in the interviewer observation and physical activity items compared to the CESD items, indicating potentially different levels of interviewer effects in different outcomes.

\begin{landscape}
\begin{table}
\begingroup\fontsize{11pt}{11pt}\selectfont
\caption{Distribution of Outcome variables in the Health and Retirement Study across Interviewers by Modes}
\label{tab:descriptiveHRS}
\begin{tabular}{lrrrrrr}
\hline
  \hline
Questions & Mean (FTF) & Mean (TEL) & \makecell{Between-\\interviewer\\ SD (FTF)} & \makecell{Between-\\interviewer\\ SD (TEL)} & \makecell{Average\\ Within-\\interviewer \\SD (FTF)} & \makecell{Average\\ Within-\\interviewer \\SD (TEL)} \\ 
\hline
\multicolumn{7}{c}{CESD questions}\\
\hline
\makecell{1. you felt depressed.} & 0.155 & 0.117 & 0.178 & 0.151 & 0.303 & 0.264 \\ 
\makecell{2. you felt that everything you did was an effort.} & 0.329 & 0.251 & 0.227 & 0.212 & 0.428 & 0.388 \\ 
 \makecell{3. your sleep was restless.} &  0.347 & 0.301 & 0.217 & 0.220 & 0.443 & 0.424 \\ 
  \makecell{4. you were happy (REVERSED CODE).} & 0.175 & 0.143 & 0.193 & 0.174 & 0.324 & 0.294 \\ 
 \makecell{5. you felt lonely.} &   0.203 & 0.153 & 0.186 & 0.159 & 0.364 & 0.319 \\ 
 \makecell{6. you enjoyed life (REVERSED CODE).} &  0.112 & 0.077 & 0.153 & 0.116 & 0.258 & 0.212 \\ 
 \makecell{7. you felt sad.} &  0.243 & 0.192 & 0.216 & 0.186 & 0.378 & 0.351 \\ 
 \makecell{8. you could not get going.} &  0.210 & 0.171 & 0.179 & 0.171 & 0.373 & 0.332 \\ 
 \makecell{9. Depressed ($\ge 4$ symptoms)} &  0.182 & 0.119 & 0.188 & 0.141 & 0.335 & 0.277 \\ 
 \hline
\multicolumn{7}{c}{Interviewer Observations}\\
\hline
  \makecell{10. attentive to the questions} & 
0.799 & 0.793 & 0.209 & 0.235 & 0.336 & 0.319 \\ 
\makecell{11. understanding of the questions} &    0.459 & 0.469 & 0.270 & 0.304 & 0.440 & 0.403 \\ 
\makecell{12. cooperation} &   0.718 & 0.663 & 0.258 & 0.281 & 0.376 & 0.396 \\ 
\makecell{13. difficulty remembering things} &  0.540 & 0.591 & 0.289 & 0.313 & 0.419 & 0.385 \\ 
 \makecell{14. difficulty hearing you} &  0.807 & 0.741 & 0.198 & 0.252 & 0.322 & 0.361 \\ 
 \makecell{15. quality of this interview} &  0.592 & 0.624 & 0.322 & 0.327 & 0.380 & 0.363 \\ 
 \hline
\multicolumn{6}{c}{Physical activity}\\
\hline
 \makecell{16. vigorous sports or activities} & 
 0.353 & 0.347 & 0.219 & 0.221 & 0.441 & 0.451 \\ 
\makecell{17. moderately energetic sports or activities} & 
  0.679 & 0.657 & 0.213 & 0.232 & 0.423 & 0.439 \\ 
\makecell{18. mildly energetic sports or activities} &   0.809 & 0.779 & 0.167 & 0.201 & 0.357 & 0.374 \\ 
   \hline
\hline
\end{tabular}
\endgroup
\end{table}
\end{landscape}

\subsubsection{Mode Effects in Means and Interviewer Variances}
Last, we discuss the modeling results presented in Table \ref{tab3:HRSResults_addCov} using Bayesian estimation. Positive mode effects in means are found in four of the nine depression items. These items are \say{felt depressed,} \say{everything was an effort,} \say{sleep was restless,} and \say{overall indicator for depression.} For example, for a female under 60 years old, who is an English-speaking Hispanic, not in a relationship, not currently employed, and with less than a high school education, participating in a FTF interview increases the probability of being classified as depressive by 8.01\%, compared to a TEL interview. Similarly, we compute 8.01\% using $\phi(\beta_0+\beta_1+\sum_s^S \gamma_sx_{si})\beta_1$, where $\phi$ is the pdf of a standard normal distribution and S is the number of covariates ($x$). Since depressive symptoms constitute sensitive information, and admitting to them might cause embarrassment for respondents, we believe that a higher level of reported depressive symptoms is closer to the truth. For the interviewer observation items, positive mode effects in means are present in three out of six items. In the FTF mode, interviewers rated respondents as more cooperative, with better hearing and overall quality of the interview, compared to the TEL mode (Table \ref{tab3:HRSResults_addCov}). Lastly, in the physical activity items, respondents tend to report engaging in mildly energetic sports more often when responding via FTF, compared to TEL.

We observe smaller interviewer variances in the substantive responses in HRS (Table \ref{tab3:HRSResults_addCov}) compared to the ABS. For depression items, the interviewer variances in FTF and TEL range from 0.002 to 0.032, corresponding to ICCs between 0.002 and 0.031. In the physical activity items, the interviewer variances range from 0.007 (ICC: 0.007) to 0.031 (ICC: 0.030). When comparing the magnitude of interviewer variances across variables, we notice larger interviewer variances for the interviewer observation items (ranging from 0.271 [ICC: 0.273] to 0.881 [ICC: 0.788]).

In terms of mode effects in interviewer variances, we find significant differences for three out of the eighteen questions examined in the HRS study, specifically one in the depression scale and two in the interviewer observation questions (Table \ref{tab3:HRSResults_addCov}). When asking participants if they felt sad, the results reveal that FTF is associated with larger interviewer variances. Additionally, interviewer variance in the FTF mode is marginally larger than in the TEL mode for the item \say{everything was an effort}. Generally, for the depression items, the interviewer variances in the FTF mode are larger than those in the TEL mode for seven out of nine items, though not always significantly. This outcome aligns with the Arab Barometer findings and may be due to interviewers approaching sensitive items differently in FTF compared to the TEL mode.

In assessing whether respondents have any difficulty remembering and hearing things, the results suggest that TEL interviewer variances are larger than FTF interviewer variances. This finding may be attributed to interviewers having fewer cues to evaluate interview quality in TEL, as opposed to FTF, where interviewers can rely on respondents' facial expressions or body language to infer participants' ability to hear questions. This might lead to responses being primarily determined by interviewers' subjective judgments and thus causing larger variances. Regarding the physical activity items, there is no evidence to reject the null hypothesis that interviewer variances are equal between modes.

It is not surprising to find higher correlations ($\rho >0.8$) between the random interviewer effects across modes for interviewer observation variables, which interviewers directly answer. In contrast, for the other two scales (CESD and physical activity scales), the effects of interviewers on responses are mediated through respondents, resulting in a smaller and less stable correlation between the FTF and TEL modes. 

Although we focus on reporting the Bayesian results, we provide the inferences from both the likelihood and the Bayesian procedures in Appendix F. We note that, in general, the estimates from the two procedures are similar, except when estimating the correlation ($\rho$). The correlations are associated with wide intervals in the CESD scales and the physical activity items. Moreover, the point estimates of the correlation are sometimes quite different between the two procedures, especially for the two types of items mentioned above. On two items, \say{happy} and \say{felt sad}, the correlation cannot be estimated using the likelihood approach. This might be due to the small interviewer variances in the scale, making the estimation of the covariance numerically challenging and thus unstable. Additionally, this might be attributed to the unbalanced interviewer burden between modes. Approximately 30\% of interviewers only conduct interviews in one mode, and 51\% of interviewers carry out fewer than five interviews in either FTF or TEL. This imbalance may result in insufficient information for estimating $\rho$.

To address the numerical challenges and evaluate whether the estimation of other parameters (e.g., $\sigma^2_f$, $\sigma^2_t$, and $\alpha$) is sensitive to $\rho$, we set $\rho$ to 0 and to the posterior mean obtained with the Bayesian procedure, and rerun Model \ref{Model:hrs_binary} for the CESD items. We find that the estimates of the interviewer variances remain nearly unchanged when specifying $\rho$ to different values or estimating $\rho$ (see details in Appendix G). Thus, we conclude that there is little sensitivity in the inferences provided by the likelihood estimation to $\rho$.

\begin{landscape}
\begin{ThreePartTable}
\begin{TableNotes}
\small
    \item Notes: $\beta_1$ is the mode effects in means, computed as the mean of the FTF estimate minus the mean of the TEL estimate. $\sigma^2_f$ is the FTF interviewer variances. $\sigma^2_t$ is the TEL interviewer variance. $\rho_{f,int}$ is the interviewer intraclass correlation associated with the FTF mode. $\rho_{t,int}$ is the interviewer intraclass correlation associated with the TEL mode. $\alpha$ refers to the log differences between the FTF and TEL interviewer variances. $\rho$ is the correlation between the FTF and TEL random interviewer effects.
\end{TableNotes}
\begingroup\fontsize{10pt}{10pt}\selectfont
\begin{longtable}{
@{}
>{\raggedright}p{0.15\linewidth}
>{\raggedright}p{0.10\linewidth}
>{\raggedright}p{0.10\linewidth}
>{\raggedright}p{0.10\linewidth}
>{\raggedright}p{0.11\linewidth}
>{\raggedright}p{0.11\linewidth}
>{\raggedright}p{0.11\linewidth}
p{0.11\linewidth}
@{}}
\caption{Interviewer Variances Per Mode for Selected Items in Health and Retirement Study Adjusting for Covariates Using Bayesian Estimation} \\ 
  \hline
\hline
Questions & $\sigma^2_f$ & $\sigma^2_t$ & $\rho_{f,int}$ & $\rho_{t,int}$ &  $\alpha$ & $\beta_1$ &$\rho$ \\
\hline
\multicolumn{8}{c}{CESD questions}\\
\hline
\makecell{felt depressed} &0.011 \hspace{1cm} [0, 0.022] & 0.013 \hspace{1cm} [0, 0.03] & 0.011 \hspace{1cm} [0.000, 0.022] & 0.013 \hspace{1cm} [0.000, 0.029] & 0.044 \hspace{1cm} [-1.148, 1.533] & \textbf{0.056} \hspace{1cm} [0.005, 0.114] & 0.07 \hspace{1.5cm} [-0.551, 0.874] \\ 
\hline
\makecell{everything was\\ an effort}  & 0.025 \hspace{1cm} [0.013, 0.037] & 0.007 \hspace{1cm} [0.001, 0.016] & 0.024 \hspace{1cm} [0.013, 0.036] & 0.007 \hspace{1cm} [0.001, 0.016] & 0.746 \hspace{1cm} [-0.002, 1.496] & \textbf{0.118} \hspace{1cm} [0.071, 0.175] & -0.128 \hspace{1.5cm} [-0.56, 0.254] \\ 
\hline
\makecell{restless sleep}  &  0.002 \hspace{1cm} [0, 0.007] & 0.005 \hspace{1cm} [0, 0.012] & 0.002 \hspace{1cm} [0.000, 0.007] & 0.005 \hspace{1cm} [0.000, 0.012] & -0.486 \hspace{1cm} [-1.89, 0.925] & \textbf{0.053} \hspace{1cm} [0.011, 0.095] & 0.337 \hspace{1.5cm} [-0.162, 0.849] \\
\hline
\makecell{happy}  & 0.011 \hspace{1cm} [0.003, 0.021] & 0.011 \hspace{1cm} [0, 0.022] & 0.011 \hspace{1cm} [0.003, 0.021] & 0.011 \hspace{1cm} [0.000, 0.022] & 0.128 \hspace{1cm} [-0.889, 1.333]] & 0.032 \hspace{1cm} [-0.024, 0.083] & \textbf{-0.518} \hspace{1.5cm} [-0.989, -0.006] \\
\hline
\makecell{lonely}  & 0.006 \hspace{1cm} [0, 0.014] & 0.006 \hspace{1cm} [0, 0.016] & 0.006 \hspace{1cm} [0.000, 0.014] & 0.006 \hspace{1cm} [0.000, 0.016] & 0.178 \hspace{1cm} [-1.455, 1.846]] & 0.048 \hspace{1cm} [-0.005, 0.099] & 0.055 \hspace{1.5cm} [-0.108, 0.218] \\ 
\hline
\makecell{enjoyed life} & 0.01 \hspace{1cm} [0.001, 0.021] & 0.007 \hspace{1cm} [0, 0.025] & 0.010 \hspace{1cm} [0.001, 0.021] & 0.007 \hspace{1cm} [0.000, 0.024] & 0.551 \hspace{1cm} [-1.096, 2.148]] & 0.061 \hspace{1cm} [-0.005, 0.134] & \textbf{0.56} \hspace{1.5cm} [0.223, 0.921] \\ 
\hline
\makecell{felt sad} & 0.032 \hspace{1cm} [0.018, 0.048] & 0.003 \hspace{1cm} [0, 0.009] & 0.031 \hspace{1cm} [0.018, 0.046] & 0.003 \hspace{1cm} [0.000, 0.009] & \textbf{1.694} \hspace{1cm} [0.463, 3.775] & 0.046 \hspace{1cm} [-0.01, 0.097] & \textbf{0.296} \hspace{1.5cm} [0.037, 0.577] \\
\hline
\makecell{could not get going} & 0.02 \hspace{1cm} [0.007, 0.029] & 0.02 \hspace{1cm} [0.006, 0.035] & 0.020 \hspace{1cm} [0.007, 0.028] & 0.020 \hspace{1cm} [0.006, 0.034] & -0.051 \hspace{1cm} [-0.732, 0.515] & 0.051 \hspace{1cm} [-0.01, 0.109] & 0.274 \hspace{1.5cm} [-0.346, 0.797] \\  
\hline
\makecell{overall indicator}  & 0.016 \hspace{1cm} [0.002, 0.024] & 0.012 \hspace{1cm} [0.001, 0.027] & 0.016 \hspace{1cm} [0.002, 0.023] & 0.012 \hspace{1cm} [0.001, 0.026] & 0.093 \hspace{1cm} [-0.901, 1.075] & \textbf{0.15} \hspace{1cm} [0.102, 0.207] & 0.244 \hspace{1.5cm} [-0.18, 0.573] \\
\hline
\multicolumn{8}{c}{Interviewer Observations}\\
\hline
\makecell{attentive}  & 0.298 \hspace{1cm} [0.233, 0.356] & 0.351 \hspace{1cm} [0.262, 0.431] & 0.230 \hspace{1cm} [0.189, 0.263] & 0.260 \hspace{1cm} [0.208, 0.301] & -0.081 \hspace{1cm} [-0.197, 0.038] & 0.018 \hspace{1cm} [-0.049, 0.088] & \textbf{0.878} \hspace{1.5cm} [0.803, 0.955] \\
\hline
\makecell{understanding} & 0.413 \hspace{1cm} [0.341, 0.493] & 0.465 \hspace{1cm} [0.366, 0.56] & 0.292 \hspace{1cm} [0.254, 0.330] & 0.317 \hspace{1cm} [0.268, 0.359] & -0.058 \hspace{1cm} [-0.149, 0.043] & 0 \hspace{1cm} [-0.064, 0.061] & \textbf{0.91} \hspace{1.5cm} [0.861, 0.958] \\ 
\hline
\makecell{cooperation} & 0.459 \hspace{1cm} [0.378, 0.556] & 0.41 \hspace{1cm} [0.321, 0.51] & 0.315 \hspace{1cm} [0.274, 0.357] & 0.291 \hspace{1cm} [0.243, 0.338] & 0.057 \hspace{1cm} [-0.039, 0.138] & \textbf{0.178} \hspace{1cm} [0.108, 0.236] & \textbf{0.931} \hspace{1.5cm} [0.881, 0.971]\\
\hline
\makecell{remembering} & 0.483 \hspace{1cm} [0.392, 0.574] & 0.605 \hspace{1cm} [0.489, 0.721] & 0.326 \hspace{1cm} [0.282, 0.365] & 0.377 \hspace{1cm} [0.328, 0.419] & \textbf{-0.112} \hspace{1cm} [-0.205, -0.028]] & -0.062 \hspace{1cm} [-0.124, 0.002] & \textbf{0.931} \hspace{1.5cm} [0.885, 0.972] \\ 
\hline
\makecell{hearing}  &0.271 \hspace{1cm} [0.212, 0.335] & 0.375 \hspace{1cm} [0.274, 0.462] & 0.213 \hspace{1cm} [0.175, 0.251] & 0.273 \hspace{1cm} [0.215, 0.316] & \textbf{-0.161} \hspace{1cm} [-0.284, -0.037] & \textbf{0.151} \hspace{1cm} [0.084, 0.229] & \textbf{0.87} \hspace{1.5cm} [0.795, 0.947]] \\ 
\hline
\makecell{Overall quality} & 0.881 \hspace{1cm} [0.749, 1.04] & 0.788 \hspace{1cm} [0.641, 0.949] & 0.468 \hspace{1cm} [0.428, 0.510] & 0.441 \hspace{1cm} [0.391, 0.487] & 0.057 \hspace{1cm} [-0.032, 0.14] & \textbf{0.086} \hspace{1cm} [0.014, 0.158] & \textbf{0.94} \hspace{1.5cm} [0.913, 0.983] \\ 
\hline
\multicolumn{8}{c}{Physical activity}\\
\hline
\makecell{vigorous sports}  & 0.017 \hspace{1cm} [0.007, 0.026] & 0.007 \hspace{1cm} [0, 0.015] & 0.017 \hspace{1cm} [0.007, 0.025] & 0.007 \hspace{1cm} [0.000, 0.015] & 0.523 \hspace{1cm} [-0.209, 1.45] & -0.037 \hspace{1cm} [-0.081, 0.014] & 0.36 \hspace{1.5cm} [-0.446, 0.827] \\ 
\hline
\makecell{moderate sport} &  0.015 \hspace{1cm} [0.006, 0.024] & 0.019 \hspace{1cm} [0.004, 0.033] & 0.015 \hspace{1cm} [0.006, 0.023] & 0.019 \hspace{1cm} [0.004, 0.032] & -0.086 \hspace{1cm} [-0.655, 0.464] & 0.031 \hspace{1cm} [-0.019, 0.078]& 0.233 \hspace{1.5cm} [-0.351, 0.698] \\ 
\hline
\makecell{mild sport}  & 0.02 \hspace{1cm} [0.002, 0.03] & 0.031 \hspace{1cm} [0.014, 0.052] & 0.020 \hspace{1cm} [0.002, 0.029] & 0.030 \hspace{1cm} [0.014, 0.049] & -0.355 \hspace{1cm} [-1.097, 0.324] & \textbf{0.134} \hspace{1cm} [0.073, 0.184] & 0.144 \hspace{1.5cm} [-0.264, 0.962] \\
\hline
\hline
 \insertTableNotes\\
\label{tab3:HRSResults_addCov}
\end{longtable}
\endgroup
\end{ThreePartTable}
\end{landscape}

\section{Simulation Study}
To understand the repeated sampling properties of our proposed method, including the power to detect mode effects in the typically modest interviewer sample sizes available, we conducted simulation studies using the ABS and the HRS setup.

\subsection{Arab Barometer Study}
This simulation study is designed such that the number of respondents ($n=2521$) and interviewers (13 in the TEL mode and 31 in the FTF mode) are the same as the ABS, as well as how respondents are matched to interviewers. We consider four scenarios, 1) no difference scenario where the FTF interviewer variance is equal to the TEL interviewer variance ($\sigma^2_f = \sigma^2_t = 0.14, \alpha_0 = -0.98$ and $\alpha = 0$), 2) small differences where $\sigma^2_f = 0.20, \sigma^2_t = 0.14, \alpha_0 = -0.98$ and $\alpha = 0.18$, 3) medium differences where $\sigma^2_f = 0.24, \sigma^2_t = 0.14, \alpha_0 = -0.98$ and $\alpha = 0.27$, and 4) large differences where $\sigma^2_f = 0.50, \sigma^2_t = 0.14, \alpha_0 = -0.98$ and $\alpha = 0.64$. We consider the true data generation model as follows:
\begin{equation*}
\begin{split}
    \eta_i= & \: \Phi(\beta_0 +\beta_1M_{ij}+b_{j(m)}),\\
    &b_{j(m)} \sim N(0,\sigma^2_{m}),\\
    &y_i \sim Bernoulli(\eta_i),
\end{split}
\end{equation*}

where $i$ indexes respondents, $j$ indexes interviewers, $m$ indicates modes ($f$ or $t$), $\Phi()$ is the cumulative distribution function of the standard normal distribution, and $M$ is a $n \times 1$ vector of the mode that each participant used to participate in the survey. 

We fit the same analytical model (\ref{Model:Arab_base}) to the simulated data, implemented separately using Proc Nlmixed and Proc MCMC in the SAS programming language. The simulation is repeated $K=200$ times, where for each iteration, the point estimates, standard errors, and 95\% confidence intervals or credible intervals of $\beta_1$, $\sigma^2_f$, $\sigma^2_t$, and $\alpha$ are computed and saved. Based on these statistics, we report the bias, coverage rate, SE ratio, and power in each scenario for the parameters.

\begin{equation*}
\begin{split}
    &Bias(\hat{\delta}) = \frac{1}{K}\sum_k^K \hat{\delta_k}-\delta,\\
    &Coverage \; Rate(\hat{\delta}) = \frac{1}{K}\sum_k^K I(\hat{\delta}_{k,lw}<\delta \, \& \, \hat{\delta}_{k,up}>\delta),\\
    &SE \; Ratio(\hat{\delta}) = \frac{1}{K}\sum_k^K \sqrt{\hat{var}(\hat{\delta_k})}/\sqrt{\frac{1}{K-1}\sum_k^K(\hat{\delta_k}-\bar{\hat{\delta_k}})^2},\\
    &Power(\hat{\delta}) = 1-\frac{1}{K}\sum_k^K I(\hat{\delta}_{k,lw}<0 \, \& \, \hat{\delta}_{k,up}>0) \; when \; \delta \neq 0,
\end{split}
\end{equation*}

where $\delta$ refers to the parameters that we are interested in estimating (i.e., $\sigma^2_f$, $\sigma^2_t$, $\beta_1$, and $\alpha$), $\hat{\delta_k}$ is the estimated point estimate of $\delta$ obtained in iteration K, $\hat{\delta}_{k,lw}$ and $\hat{\delta}_{k,up}$ is the lower bound and upper bound of the estimated parameter.

Table \ref{tab:sim_arab} displays the simulation results using the Arab Barometer setup. When $\alpha = 0$, the power reported in Table \ref{tab:sim_arab} represents the Type 1 error rate. We observe that the power to reject the null hypothesis stating that interviewer variances are equal ($\alpha=0$) is limited across the scenarios. However, as the differences grow larger (0.18-0.64), the power does increase from 0.075 to 0.520 in the Bayesian procedure and from 0.110 to 0.633 in the frequentist approach. There are some differences in the power provided by the likelihood and Bayesian approaches. This is because the likelihood procedures do not offer nominal coverage rates in Scenarios 1 to 3; as a result, the power obtained from the likelihood and Bayesian procedures is based on different significance levels. The small power of $\alpha$ is primarily due to the very limited number of interviewers in both FTF and TEL modes. Conversely, the power of rejecting the null hypothesis that there are no mode effects in means ($\beta_1$) when the alternative hypothesis is true is considerably higher (around 0.90). However, as $\alpha$ becomes larger and the interviewer variances increase simultaneously, we observe a declining power of $\beta_1$, due to the decline in effective sample size from the increased ICC.

\begin{ThreePartTable}
\begingroup\fontsize{11pt}{11pt}\selectfont
\begin{TableNotes}
\small
    \item Notes: $\beta_1$ is the mode effects in means, computed as the mean of the FTF estimate minus the mean of the TEL estimate. $\sigma^2_f$ is the FTF interviewer variances. $\sigma^2_t$ is the TEL interviewer variance. $\alpha$ refers to the log differences between the FTF and TEL interviewer variances. 
\end{TableNotes}
\begin{longtable}{
@{}
>{\raggedleft}p{0.10\linewidth}
>{\raggedleft}p{0.09\linewidth}
>{\raggedright}p{0.09\linewidth}
>{\raggedright}p{0.09\linewidth}
>{\raggedright}p{0.09\linewidth}
>{\raggedleft}p{0.09\linewidth}
>{\raggedright}p{0.09\linewidth}
>{\raggedright}p{0.09\linewidth}
p{0.09\linewidth}
@{}}
\caption{Simulation study using the Arab Barometer Setup} \\ 
\hline
  \hline
\multirow{2}{*}{Parameters} & \multicolumn{4}{c}{Likelihood results} & \multicolumn{4}{c}{Bayesian results} \\
  \cmidrule(lr){2-5}
  \cmidrule(lr){6-9}
& Bias & Coverage rate & SE ratio & Power & Bias & Coverage rate & SE ratio & Power\\ 
\hline

\multicolumn{9}{c}{Scenario 1: No differences}\\
\hline  
  $\sigma^2_f=0.14$ & -0.002 & 0.950 & 1.000 & N/A & 0.017 & 0.940 & 1.059 & N/A \\ 
  $\sigma^2_t=0.14$ & -0.001 & 0.955 & 1.014 & N/A & 0.049 & 0.975 & 1.346 & N/A \\
  $\beta_1=0.5$ & -0.003 & 0.965 & 1.023 & 0.935 & 0.006 & 0.955 & 1.121 &  0.930\\
  $\alpha=0$ & 0.028 & 0.930 & 0.888 & 0.070 & -0.033 & 0.985 & 1.107 &  0.015\\ 
\hline
\multicolumn{9}{c}{Scenario 2: Small differences}\\
\hline 
  $\sigma^2_f=0.20$ &-0.012 & 0.960 & 0.948 & N/A & 0.028 & 0.975 & 1.105 &  N/A\\ 
  $\sigma^2_t=0.14$ & -0.007 & 0.935 & 0.974 & N/A & 0.059 & 0.955 & 1.161 &  N/A\\ 
  $\beta_1=0.5$ & -0.002 & 0.940 & 0.926 & 0.950 & -0.001 & 0.950 & 1.078 & 0.900  \\ 
  $\alpha=0.18$ & 0.042 & 0.920 & 0.928 & 0.110 & -0.020 & 0.950 & 0.955 & 0.075  \\ 
\hline
\multicolumn{9}{c}{Scenario 3: Medium differences}\\
\hline 
  $\sigma^2_f=0.24$ &-0.002 & 0.920 & 0.947 & N/A & 0.039 & 0.920 & 0.980 & N/A\\ 
  $\sigma^2_t=0.14$ & -0.013 & 0.955 & 1.009 & N/A & 0.061 & 0.980 & 1.311 & N/A \\ 
  $\beta_1=0.5$ & 0.004 & 0.935 & 0.940 & 0.920 & -0.010 & 0.960 & 1.184 & 0.860  \\ 
  $\alpha=0.27$ & 0.079 & 0.905 & 0.922 & 0.230 & -0.042 & 0.960 & 1.075 & 0.085  \\ 
\hline
\multicolumn{9}{c}{Scenario 4: Large differences}\\
\hline  
  $\sigma^2_f=0.50$ &-0.007 & 0.970 & 1.058 & N/A & 0.078 & 0.950 & 1.093 & N/A\\ 
  $\sigma^2_t=0.14$ & -0.009 & 0.960 & 1.055 & N/A & 0.054 & 0.935 & 1.231 & N/A \\ 
  $\beta_1=0.5$ & 0.022 & 0.935 & 0.965 & 0.824 & -0.016 & 0.980 & 1.097 & 0.690 \\ 
  $\alpha=0.64$ & 0.079 & 0.945 & 0.906 & 0.633 & 0.012 & 0.955 & 0.882 & 0.520  \\ 
\hline
\insertTableNotes\\
\label{tab:sim_arab}
\end{longtable}
\endgroup
\end{ThreePartTable}

\subsection{Health and Retirement Study}
In the simulation study using the HRS setup, we consider the following data generation model using the same notations as in the ABS simulation study. We use $b_{jf}$ to represent random interviewer effects in the FTF mode and $b_{jt}$ to represent random interviewer effects in the TEL mode:
\begin{equation*}
\begin{split}
    & \eta_i=  \: \Phi(\beta_0 +\beta_1M_{ij}+b_{jf} M_{ij} + b_{jt}(1-M_{ij})),\\
     & \begin{pmatrix}
         b_{jf} \\ b_{jt}
        \end{pmatrix} \sim N \Bigg( \begin{pmatrix}
        0 \\ 0
        \end{pmatrix}, \begin{pmatrix}
        \sigma^2_{f} & \rho \sigma_{f}  \sigma_{t} \\
        \rho \sigma_{f}\sigma_{t} & \sigma^2_{t} 
        \end{pmatrix} \Bigg).\\ 
    &y_i \sim Bernoulli(\eta_i),
\end{split}
\end{equation*}

We consider four scenarios: 1) $\sigma^2_f = \sigma^2_t=0.03, \alpha_0 = -1.75$, and $\alpha = 0$; 2) $\sigma^2_f = 0.05, \sigma^2_t=0.03, \alpha_0 = -1.75$, and $\alpha = 0.26$; 3) $\sigma^2_f = 0.06, \sigma^2_t=0.03, \alpha_0 = -1.75$, and $\alpha = 0.35$; 4) $\sigma^2_f = 0.09, \sigma^2_t=0.03, \alpha_0 = -1.75$, and $\alpha = 0.55$. Across all scenarios, $\beta_1 = 0.5$ and $\rho =0.5$. We report bias, coverage rate, SE ratio, power for these parameters and the logarithmic differences of interviewer variances between FTF and TEL ($\alpha$) in Table \ref{tab:sim_HRS}.

Table \ref{tab:sim_HRS} illustrates that as $\alpha$ rises from 0 to 0.55, the power correspondingly increases from 0.035 to 0.990 using the Bayesian procedure, and from 0.035 to 0.935 employing the likelihood approach. The findings suggest that When $\alpha$ is large enough, we can achieve a reasonably high power using the HRS setup. Upon comparing Table \ref{tab:sim_arab} and Table \ref{tab:sim_HRS}, we observe that the power to reject the null hypothesis asserting equal interviewer variances, when the alternative hypothesis holds true, surpasses that in the ABS simulation. This outcome aligns with expectations, given the larger number of interviewers involved in the HRS. In addition, we note that the likelihood approach may not always reach the 95\% nominal coverage rates (in Scenarios 3 and 4), thus the power computed using the likelihood and the Bayesian procedures are based on different significance levels.

\begin{ThreePartTable}
   \label{tab:sim_HRS}
    \begin{TableNotes}
    \small
        \item Notes: $\beta_1$ is the mode effects in means, computed as the mean of the FTF estimate minus the mean of the TEL estimate. $\sigma^2_f$ is the FTF interviewer variances. $\sigma^2_t$ is the TEL interviewer variance. $\alpha$ refers to the log differences between the FTF and TEL interviewer variances. $\rho$ is the correlation between the FTF and TEL random interviewer effects.
    \end{TableNotes}
    \begingroup\fontsize{11pt}{11pt}\selectfont
    \begin{longtable}{
    @{}
    >{\raggedright}p{0.09\linewidth}
    >{\raggedright}p{0.09\linewidth}
    >{\raggedright}p{0.09\linewidth}
    >{\raggedright}p{0.09\linewidth}
    >{\raggedright}p{0.09\linewidth}
    >{\raggedright}p{0.09\linewidth}
    >{\raggedright}p{0.09\linewidth}
    >{\raggedright}p{0.09\linewidth}
    p{0.09\linewidth}
    @{}}
    \caption{Simulation study using the HRS Setup} \\ 
    \hline
      \hline
    \multirow{2}{*}{Parameters} & \multicolumn{4}{c}{Likelihood results} & \multicolumn{4}{c}{Bayesian results} \\
      \cmidrule(lr){2-5}
      \cmidrule(lr){6-9}
    & Bias & Coverage rate & SE ratio & Power & Bias & Coverage rate & SE ratio & Power\\ 
      \hline
      \multicolumn{9}{c}{Scenario 1: No differences}\\
    \hline
      $\sigma_f^2=0.03$ & -0.000 & 0.980 & 1.085 & N/A & 0.003 & 0.965 & 1.704 & N/A\\ 
      $\sigma_t^2=0.03$ & -0.001 & 0.975 & 1.049 & N/A & 0.002 & 0.935 & 1.469 & N/A\\ 
      $\beta_1=0.5$ & -0.002 & 0.940 & 1.029 & 1.000 & -0.000 & 0.960 & 1.128 & 1.000\\ 
      $\rho=0.5$ &0.012 & 0.965 & 1.009 & 0.470 & -0.020 & 0.925 & 1.061 &  0.690\\ 
      $\alpha=0$ &  0.022 & 0.965 & 1.019 & 0.035 & 0.047 & 0.965 & 0.928 & 0.035\\ 
       \hline
    \multicolumn{9}{c}{Scenario 2: Small differences}\\
    \hline
      $\sigma_f^2=0.05$ & 0.000 & 0.940 & 0.999 & N/A & 0.001 & 0.955 & 1.507 & N/A\\ 
      $\sigma_t^2=0.03$ & -0.000 & 0.975 & 1.125 & N/A & 0.002 & 0.945 & 1.249 & N/A\\ 
      $\beta_1=0.5$ & 0.003 & 0.960 & 0.996 & 1.000 & 0.001 & 0.950 & 0.983 & 1.000 \\
      $\rho=0.5$ &0.020 & 0.980 & 1.084 & 0.695 & -0.021 & 0.925 & 1.032 & 0.755 \\ 
      $\alpha=0.26$ &  0.018 & 0.940 & 0.978 & 0.270 & 0.008 & 0.940 & 0.934 & 0.295\\
    \hline
    \multicolumn{9}{c}{Scenario 3: Medium differences}\\
    \hline 
      $\sigma_f^2=0.06$ & -0.001 & 0.945 & 0.999 & N/A & 0.001 & 0.950 & 1.268 & N/A \\ 
      $\sigma_t^2=0.03$ & -0.001 & 0.975 & 1.045 & N/A & 0.002 & 0.940 & 1.103 & N/A\\ 
      $\beta_1=0.5$ & -0.001 & 0.920 & 0.993 & 1.000 & 0.001 & 0.965 & 1.007 & 1.000 \\
      $\rho=0.5$ &0.011 & 0.970 & 1.030 & 0.665 & -0.009 & 0.930 & 1.014 & 0.815 \\ 
      $\alpha=0.35$ & 0.024 & 0.910 & 0.919 & 0.510 & 0.008 & 0.945 & 0.949 & 0.530 \\
      \hline
    \multicolumn{9}{c}{Scenario 4: Large differences}\\
    \hline  
      $\sigma_f^2=0.09$ & 0.000 & 0.930 & 0.983 & N/A & 0.002 & 0.950 & 1.201 & N/A \\ 
      $\sigma_t^2=0.03$ & -0.001 & 0.955 & 1.054 & N/A & -0.001 & 0.915 & 1.089 & N/A\\ 
      $\beta_1=0.5$ & 0.004 & 0.950 & 1.009 & 1.000 & -0.002 & 0.955 & 1.031 & 1.000 \\
      $\rho=0.5$ &0.009 & 0.985 & 1.085 & 0.750 & 0.004 & 0.970 & 1.121 & 0.860 \\ 
      $\alpha=0.55$ & 0.029 & 0.915 & 0.977 & 0.935 & 0.070 & 0.950 & 0.955 & 0.990 \\
    \hline
    \hline
     \insertTableNotes\\
    \end{longtable}
    \endgroup 
\end{ThreePartTable}

\section{Discussion}
This paper explores the presence of mode effects in interviewer variances across multiple items in two national surveys. In the ABS, we find statistical evidence for differing interviewer effects between the FTF and TEL modes in one (marginally) out of six sensitive items and one out of three item missing indicators. Besides, for sensitive items and missing indicators in the ABS, interviewer variances from the FTF mode are generally larger than those from the TEL mode. Meanwhile, we should interpret the ABS results with caution. Due to the small number of interviewers used in the study, null findings cannot be translated into small or no effects, somewhat hampering the strength of the evidence. Utilizing the 2016 HRS data, we observe significant mode effects in interviewer variances on two depression items (one marginally) and two interviewer observation item. For sensitive depression items, a similar pattern emerges, with larger interviewer variances in FTF than in TEL. These findings indicate that sensitive questions and item missing items are crucial challenges when stabilizing interviewer variances between modes. In addition, the magnitude of interviewer variances are much larger on interviewer observation items than substantive responses. Evidence suggests that TEL interviewer variances are larger than FTF interviewer variances on these items. This could be because these questions involve more subjective evaluations and may offer greater opportunities to reduce interviewer variances by standardizing interviewer protocols for such items, especially in the TEL mode. 

Simulation studies suggest that it is possible to achieve reasonable power with either the ABS or HRS setup if there are substantial mode effects in interviewer variances. However, with small mode effects, the power is limited, especially in the ABS setup. The observation of significant mode effects in interviewer variances in both the ABS and HRS data highlights the importance of considering the role of modes on interviewer effects, particularly when addressing sensitive topics and item nonresponse. Given the typically limited number of interviewers employed in most surveys, a null finding may not necessarily indicate equal interviewer variance. However, it is still useful for survey agencies to consider such investigation as a positive finding is valid and should capture the attention of researchers. Moreover, in the presence of multiple underpowered studies that employ few interviewers, a meta-analysis can be conducted to combine the inferences made from these studies and better explore the mode effects in interviewer variances.

The literature has extensively documented whether modes affect measurement errors at the respondent level \cite{tourangeau1996asking,kreuter2008social}. However, few studies have investigated whether and how modes influence interviewer-related measurement errors, particularly following the widespread adoption of mixed-mode designs. This paper addresses this gap by analyzing two national surveys with distinct mixed-mode design features, such as the number of interviewers and whether the interviewers are nested under modes. When interviewers are nested under modes, it is hard to determine if the observed differences are attributable to modes or interviewers. The current modeling approach presumes that all systematic differences between responses collected in TEL and FTF are a consequence of modes, not interviewers. If survey organizations possess information on interviewer characteristics, they can evaluate this assumption by comparing the characteristics of interviewers between modes. Such an analysis would help disentangle the effects of modes from those of interviewers, providing valuable insights for survey data quality. 

For designs that allow interviewers to collect data in both modes, the models presented in this paper enable the estimation of individual interviewer effects in each mode. This is useful for detecting interviewers with a substantial impact on responses in one or both modes. Utilizing these estimated interviewer effects, we can further identify if specific interviewers consistently exhibit large effects across variables, potentially signaling the need for intervention by interviewer supervisors. If particular variables are associated with significant interviewer variances in a certain mode, this may warrant improved interviewer training for those items. For instance, based on this study's findings, a more standardized interview protocol could be considered for sensitive items and when respondents answer \say{don't know} to questions in FTF mode. As such, we recommend that survey agencies incorporate these analyses into their routine data quality assessments. Future research could investigate whether interviewer characteristics can explain the differential interviewer effects observed across modes, potentially shedding light on the underlying mechanisms at play.

When determining which mode to use for generating population estimates in mixed-mode studies, it is desirable to have smaller bias and lower interviewer variances, which might result in smaller mean squared error. However, in reality, the mode with smaller bias and lower interviewer variance may not always be the same, as shown in this paper. For instance, FTF interviews may be linked with less bias but larger interviewer variance. How to balance the trade-offs between bias and variance in a formal method will be a topic for future research. This study showcases two survey examples to evaluate mode effects both in means and interviewer variances. If such analyses are routinely adopted by researchers who design and implement mixed-mode studies, more evidence can be accumulated about whether and how interviewers could have performed differently in different modes of data collection. This can become the basis for developing future mixed-mode protocols. When reporting the results of the analysis, we recommend that survey agencies explain how their interviewers are assigned to or self-select different modes and clarify whether the resultant mode effects in interviewer variances are consistent with their expectations.

In this paper, we observe some discrepancies between the results obtained from the maximum likelihood procedure and the Bayesian procedure implemented in the SAS programming language. When interviewer variances are small, fitting the analytical model with correlated interviewer random effects across modes using the likelihood approach can be challenging. In this situation, the Bayesian approach can be particularly useful, as employing proper and informative priors helps ensure that we draw inferences from proper posterior distributions.

This study has three main limitations. First, like other similar studies \cite{west2022interviewer,groves1986measuring}, it faces the issue of limited statistical power, as demonstrated in the simulation study. Second, we consider dichotomized outcomes in this study due to computational reasons; however, this may not be an optimal approach for studying interviewer variance, as collapsing categories may reduce variances. Future studies can explore this research question using different types of outcomes and larger sample sizes. Last, both surveys lack randomization in the interviewer assignment scheme. Ideally, when estimating interviewer variances, interpenetrated designs should be used to ensure that the variability is solely due to the interviewer measurement process, rather than differences among respondents. As a workaround for the absence of randomization, we included respondent characteristics in the analysis model. However, interviewer variances might still be overestimated due to unobserved covariates not accounted for in the models.


\subsection*{Acknowledgements}

 This project was supported by the Daniel Katz Dissertation Fellowship in Psychology and Survey Methodology at the University of Michigan Institute for Social Research. The authors thank the investigators, the staff, and the participants of the Arab Barometer Study and the Health and Retirement study for their valuable contributions. We appreciate Dr. Brady West for his help in obtaining the Health and Retirement Study interviewer data. We acknowledge that we utilized the AI language model, ChatGPT, developed by OpenAI, for assistance with grammar correction and refinement of the wording in this paper.

\newpage
\setcounter{table}{0}
\renewcommand{\thetable}{A\arabic{table}}

\setcounter{figure}{0}
\renewcommand{\thefigure}{A\arabic{figure}}

\subsection*{Appendix A: Outcome Variables Used in the Arab Barometer Study}
\label{Append:ArabBarometerOutcomes}

\begin{ThreePartTable}
\begingroup\fontsize{12pt}{12pt}\selectfont
\begin{longtable}{l  c  l }
\caption{Outcome Variables Used in the Arab Barometer Study} \\ 
  \hline
\hline
Questions & Original response categories & Collapsed response categories \\ 
  \hline
\hline
\multicolumn{3}{c}{Sensitive political questions}\\
\hline
\makecell{Freedom of the media to \\criticize the things\\ government does?} & \makecell{1. Guaranteed to a great extent\\ 2. Guaranteed to a medium extent\\3. Guaranteed to a limited extent\\4. Not guaranteed at all} 
 & \makecell{1. Guaranteed to a great \\
 \quad or medium extent\\
0. Guaranteed to a limited extent \\
\quad or not guaranteed at all}\\ 
\hline
\makecell{How much trust do\\ you have in government?} & \makecell{1. A great deal of trust\\ 2. Quite a lot of trust\\3. Not a lot of trust\\4. No trust at all}  & \makecell{1. A great deal of or  \\
 \quad quite a lot of trust\\
0. Not a lot of trust or \\
\quad no trust at all}  \\ 
\hline
\makecell{How much trust do\\ you have in courts \\ and the legal system?} & \makecell{1. A great deal of trust\\ 2. Quite a lot of trust\\3. Not a lot of trust\\4. No trust at all}  & \makecell{1. A great deal of or  \\
 \quad quite a lot of trust\\
0. Not a lot of trust or \\
\quad no trust at all}  \\ 
\hline
\makecell{How satisfied are you \\with the healthcare system \\in our country?} & \makecell{1. Completely satisfied\\
2. Satisfied\\
3. Dissatisfied\\
4. Completely dissatisfied}  & \makecell{1. Completely satisfied \\
\quad or satisfied\\
0. Dissatisfied or \\
\quad completely dissatisfied
}  \\ 
\hline
\makecell{How would you evaluate \\the current government’s \\performance on keeping \\prices down?} & \makecell{1. Very good\\
2. Good\\
3. Bad\\
4. Very bad
}  & \makecell{1. Very good or good\\
0. Bad or very bad
}  \\ 
\hline
\makecell{How would you evaluate \\the current government’s \\performance on responding \\to the COVID-19 outbreak?} & \makecell{1. Very good\\
2. Good\\
3. Bad\\
4. Very bad
}  & \makecell{1. Very good or good\\
0. Bad or very bad
}  \\ 
\hline
\hline
\multicolumn{3}{c}{Less Sensitive International Questions}\\
\hline
\makecell{Please tell me if \\you have a very favorable,\\ somewhat favorable, \\somewhat unfavorable, \\or very unfavorable \\opinion of the United States.} & \makecell{1. Very favorable\\
2. Somewhat favorable\\
3. Somewhat unfavorable\\
4. Very unfavorable
}  & \makecell{1. Very or somewhat \\
\quad favorable\\
0. Somewhat or very \\
\quad unfavorable
}  \\ 
\hline
\makecell{Please tell me if \\you have a very favorable,\\ somewhat favorable, \\somewhat unfavorable, \\or very unfavorable \\opinion of Germany.} & \makecell{1. Very favorable\\
2. Somewhat favorable\\
3. Somewhat unfavorable\\
4. Very unfavorable
}  & \makecell{1. Very or somewhat \\
\quad favorable\\
0. Somewhat or very \\
\quad unfavorable
}  \\ 
\hline
\makecell{Please tell me if \\you have a very favorable,\\ somewhat favorable, \\somewhat unfavorable, \\or very unfavorable \\opinion of China.} & \makecell{1. Very favorable\\
2. Somewhat favorable\\
3. Somewhat unfavorable\\
4. Very unfavorable
}  & \makecell{1. Very or somewhat \\
\quad favorable\\
0. Somewhat or very \\
\quad unfavorable
}  \\ 
\hline
\hline
\multicolumn{3}{c}{Whether missing on international questions (constructed)}\\
\hline
\makecell{Please tell me if \\you have a very favorable,\\ somewhat favorable, \\somewhat unfavorable, \\or very unfavorable \\opinion of the United States.} & \makecell{Don’t know or \\refused to answer \\(Interviewer: do not read)
}  & \makecell{1. Don’t know or \\
\quad refused to answer\\
0. Answered
}  \\ 
\hline
\makecell{Please tell me if \\you have a very favorable,\\ somewhat favorable, \\somewhat unfavorable, \\or very unfavorable \\opinion of Germany.} & \makecell{Don’t know or \\refused to answer \\(Interviewer: do not read)
}  & \makecell{1. Don’t know or \\
\quad refused to answer\\
0. Answered
}  \\ 
\hline
\makecell{Please tell me if \\you have a very favorable,\\ somewhat favorable, \\somewhat unfavorable, \\or very unfavorable \\opinion of China.} & \makecell{Don’t know or \\refused to answer \\(Interviewer: do not read)
}  & \makecell{1. Don’t know or \\
\quad refused to answer\\
0. Answered
}  \\ 
\hline
\hline
\label{tab1:variable_info}
\end{longtable}
\endgroup
\end{ThreePartTable}

\subsection*{Appendix B: Derivations of the Variance of $\alpha$ Using Delta Method}
\label{Append:Alpha}
\begin{equation*}
\begin{split}
    &var(\alpha) = var(log(\sigma_f)-log(\sigma_t))\\
                 & = var(log(\sigma_f)) + var(log(\sigma_t)) - 2cov(log(\sigma_f),log(\sigma_t))\\
                 & = \frac{1}{4}var(2log(\sigma_f)) + \frac{1}{4}var(2log(\sigma_t)) -  2cov(log(\sigma_f),log(\sigma_t))\\
                 & = \frac{1}{4}var(log(\sigma^2_f)) + \frac{1}{4}var(log(\sigma^2_t)) -  \frac{1}{4}\times2cov(log(\sigma^2_f),log(\sigma^2_t))\\
                 & = \frac{1}{4}var(log(\sigma^2_f)) + \frac{1}{4}var(log(\sigma^2_t))-\frac{1}{2}cov(log(\sigma^2_f),log(\sigma^2_t))\\
\end{split}
\end{equation*}

We express $var(\alpha)$ as a function of $var(log(\sigma^2_f))$, $var(log(\sigma^2_t))$, and $cov(log(\sigma^2_f),log(\sigma^2_t))$, as we apply a log transformation to $\sigma^2_t$ and $\sigma^2_f$ to stabilize their variances. The covariance between $log(\sigma^2_f)$ and $log(\sigma^2_t)$ can be assumed to be 0 when the random interviewer effects of FTF and TEL are not correlated, as is the case in the ABS. In contrast, in the HRS, when the random interviewer effects are correlated across modes, the covariance between the two estimates should be considered when calculating $var(\alpha)$.

\subsection*{Appendix C: Distribution of Outcome Variables per Interviewers in the Arab Barometer Study}

\begin{figure}
    \begin{subfigure}{\textwidth}
    \centering
    \includegraphics[width=0.7\textwidth]{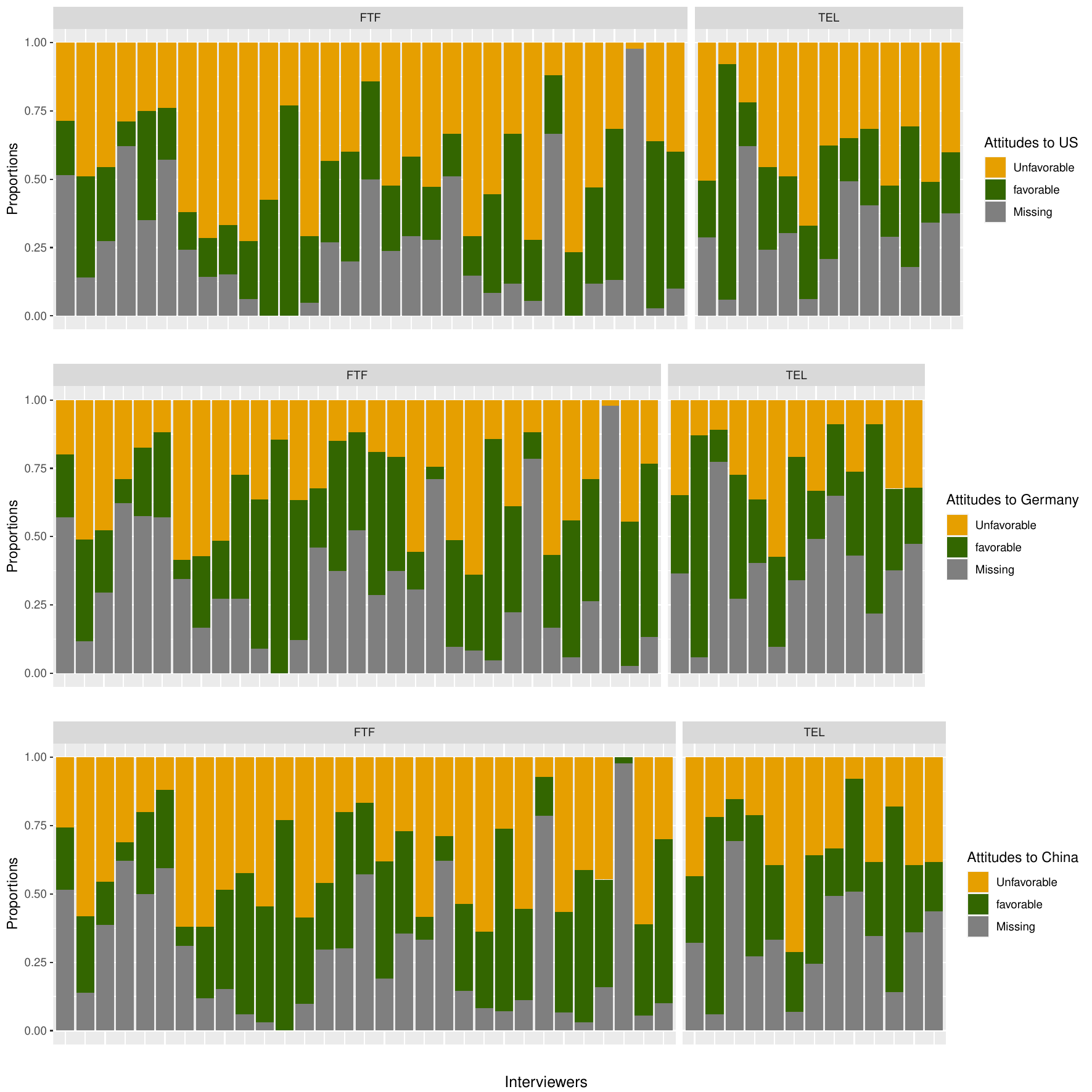}
    \caption{International Attitude Outcomes}
    \end{subfigure}

\end{figure}

\begin{figure}\ContinuedFloat
    \begin{subfigure}{\textwidth}
    \centering
    \includegraphics[width=0.7\textwidth]{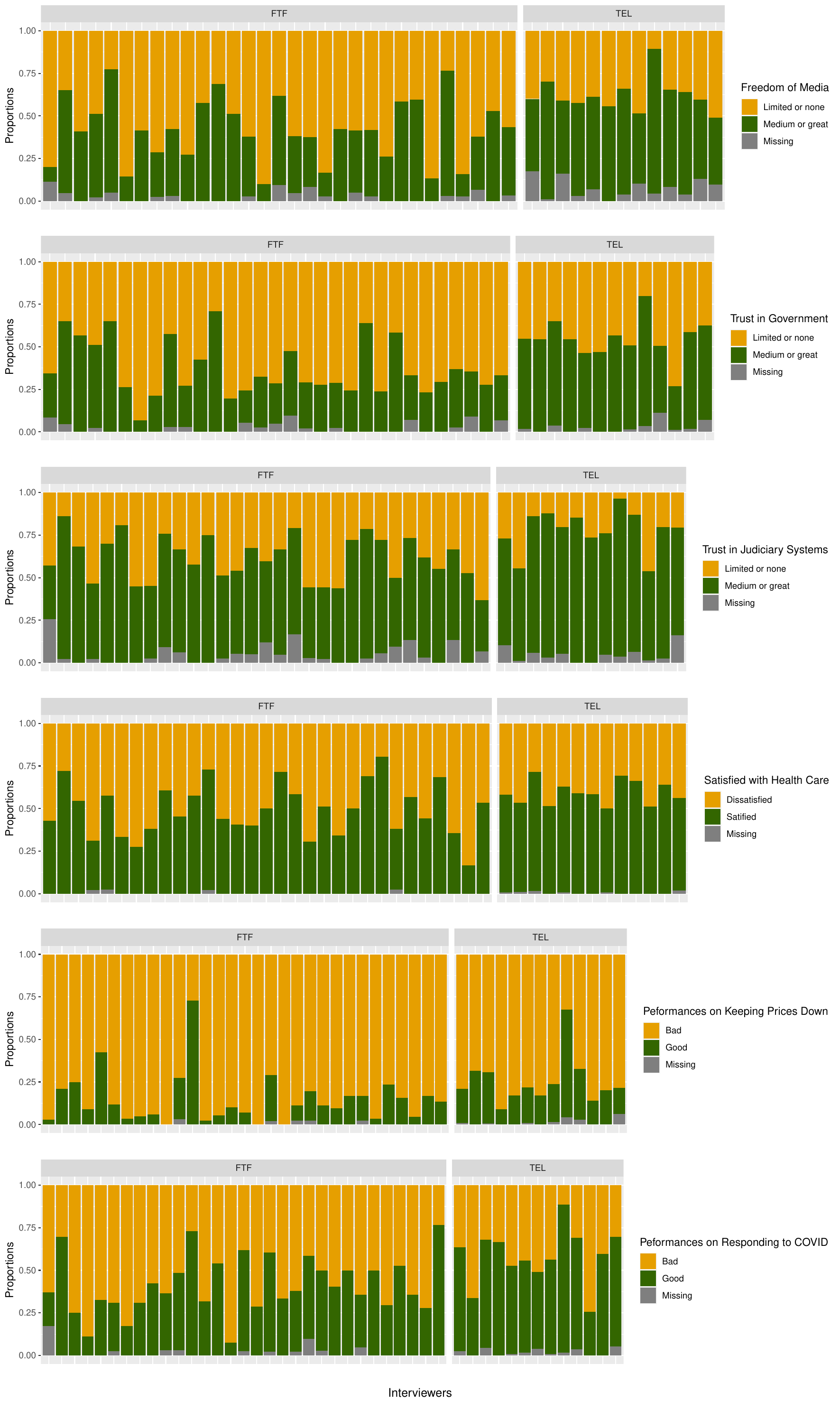}
    
    \caption{Political Outcomes}
    \end{subfigure}
    \caption{Distribution of Outcome Variables in the Arab Barometer Study}
    \label{fig:distOutbyInt}
\end{figure}

\clearpage

\subsection*{Appendix D: Full Results on Interviewer Variances in the Arab Barometer Study}
\begin{landscape}
\begin{ThreePartTable}
\begin{TableNotes}
\small
    \item Notes: $\beta_1$ is the mode effects in means, computed as the mean of the FTF estimate minus the mean of the TEL estimate. $\sigma^2_f$ is the FTF interviewer variances. $\sigma^2_t$ is the TEL interviewer variance. $\rho_{f,int}$ is the interviewer intraclass correlation associated with the FTF mode. $\rho_{t,int}$ is the interviewer intraclass correlation associated with the TEL mode. $\alpha$ refers to the log differences between the FTF and TEL interviewer variances.
\end{TableNotes}
\begingroup\fontsize{11pt}{11pt}\selectfont
\begin{longtable}{
@{}
>{\raggedright}p{0.09\linewidth}
>{\raggedright}p{0.09\linewidth}
>{\raggedright}p{0.09\linewidth}
>{\raggedright}p{0.09\linewidth}
>{\raggedright}p{0.09\linewidth}
>{\raggedright}p{0.09\linewidth}
>{\raggedright}p{0.09\linewidth}
>{\raggedright}p{0.09\linewidth}
p{0.09\linewidth}
@{}}
\caption{Interviewer Variances Per Mode for Selected Items in the Arab Barometer Study Adjusting for Covariates} \\ 
  \hline
\hline
\multirow{2}{*}{Questions} & \multicolumn{4}{c}{Likelihood Results} & \multicolumn{4}{c}{Bayesian Results} \\ 
  \cmidrule(lr){2-9}
 & $\sigma^2_f$ & $\sigma^2_t$ & $\alpha$ & $\beta_1$ & $\sigma^2_f$ & $\sigma^2_t$ & $\alpha$ & $\beta_1$ \\
\hline
\hline
\multicolumn{9}{c}{Sensitive political questions}\\
\hline
\makecell{1. Freedom \\of the\\ media}& 0.222 (0.071) [0.116, 0.425] & 0.092 (0.044) [0.034, 0.244] & 0.443 (0.291) [-0.127, 1.013] & \textbf{-0.529 (0.132) [-0.794, -0.263]} & 0.252 (0.082) [0.122, 0.428] & 0.135 (0.076) [0.036, 0.284] & 0.355 (0.298) [-0.223, 0.898] & \textbf{-0.526 (0.146) [-0.795, -0.222]} \\
\hline
\makecell{2. trust \\in \\government} & 0.159 (0.054) [0.081, 0.313] & 0.085 (0.041) [0.032, 0.225] & 0.311 (0.293) [-0.263, 0.886] & \textbf{-0.5 (0.121) [-0.745, -0.255]} & 0.188 (0.066) [0.083, 0.322] & 0.127 (0.079) [0.029, 0.275] & 0.239 (0.311) [-0.382, 0.838] & \textbf{-0.504 (0.134) [-0.768, -0.238]} \\  
\hline
\makecell{3. trust \\in courts} & 0.093 (0.036) [0.042, 0.202] & 0.145 (0.069) [0.056, 0.378] & -0.225 (0.306) [-0.825, 0.374] & \textbf{-0.553 (0.133) [-0.822, -0.285]} & 0.113 (0.046) [0.038, 0.201] & 0.214 (0.119) [0.05, 0.445] & -0.291 (0.327) [-0.94, 0.318] & \textbf{-0.555 (0.154) [-0.881, -0.273]} \\
\hline
\makecell{4. satisfied \\with \\healthcare} & 0.118 (0.042) [0.057, 0.244] & 0.018 (0.013) [0.004, 0.078] & \textbf{0.955 (0.411) [0.149, 1.761]} & \textbf{-0.285 (0.089) [-0.465, -0.105]} & 0.143 (0.056) [0.051, 0.251] & 0.03 (0.024) [0, 0.075] & 0.906 (0.467) [-0.054, 1.758] & \textbf{-0.278 (0.1) [-0.475, -0.085]} \\ 
\hline
\makecell{5. performance\\ on inflation} & 0.327 (0.112) [0.164, 0.653] & 0.139 (0.064) [0.055, 0.351] & 0.428 (0.286) [-0.134, 0.989] & \textbf{-0.515 (0.163) [-0.843, -0.186]} & 0.393 (0.146) [0.153, 0.672] & 0.204 (0.112) [0.051, 0.435] & 0.361 (0.305) [-0.275, 0.927] & \textbf{-0.523 (0.182) [-0.861, -0.153]} \\
\hline
\makecell{6. performance\\ during \\COVID-19} & 0.172 (0.057) [0.088, 0.336] & 0.165 (0.074) [0.067, 0.406] & 0.019 (0.278) [-0.526, 0.563] & \textbf{-0.495 (0.146) [-0.79, -0.201]} & 0.202 (0.069) [0.084, 0.34] & 0.224 (0.112) [0.07, 0.443] & -0.026 (0.285) [-0.602, 0.508] & \textbf{-0.51 (0.166) [-0.841, -0.205]} \\
\hline
\hline
\multicolumn{9}{c}{International Questions}\\
\hline
\makecell{7. favorable \\of the \\United \\States} & 0.163 (0.058) [0.08, 0.333] & 0.262 (0.115) [0.108, 0.634] & -0.238 (0.282) [-0.79, 0.314] & -0.05 (0.173) [-0.399, 0.299] & 0.198 (0.073) [0.074, 0.34] & 0.362 (0.184) [0.104, 0.719] & -0.278 (0.293) [-0.841, 0.282] & -0.057 (0.193) [-0.45, 0.318] \\ 
\hline
\makecell{8. favorable \\of Germany} & 0.245 (0.084) [0.122, 0.49] & 0.237 (0.107) [0.096, 0.588] & 0.015 (0.283) [-0.539, 0.57] & -0.142 (0.178) [-0.5, 0.216] & 0.292 (0.11) [0.12, 0.514] & 0.33 (0.164) [0.092, 0.663] & -0.037 (0.314) [-0.603, 0.548] & -0.147 (0.201) [-0.551, 0.236] \\ 
\hline
\makecell{9. favorable \\of China} & 0.174 (0.064) [0.083, 0.365] & 0.27 (0.118) [0.112, 0.654] & -0.221 (0.286) [-0.781, 0.339] & -0.142 (0.177) [-0.498, 0.214] & 0.205 (0.077) [0.083, 0.361] & 0.378 (0.196) [0.116, 0.787] & -0.282 (0.293) [-0.869, 0.245] & -0.15 (0.19) [-0.549, 0.19] \\
\hline
\hline
\multicolumn{9}{c}{Whether missing on international questions (constructed)}\\
\hline
\makecell{10. if \\missing\\ on \\ favorable \\of the \\United \\States} & 0.847 (0.264) [0.451, 1.589] & 0.252 (0.109) [0.105, 0.605] & \textbf{0.606 (0.267) [0.082, 1.13]} & -0.294 (0.228) [-0.753, 0.166] & 0.995 (0.342) [0.48, 1.71] & 0.343 (0.169) [0.104, 0.668] & \textbf{0.557 (0.279) [0.014, 1.121]} & -0.298 (0.255) [-0.805, 0.172] \\ 
\hline
\makecell{11. if \\missing\\ on \\ favorable \\of Germany} & 0.753 (0.223) [0.414, 1.37] & 0.357 (0.151) [0.153, 0.836] & 0.373 (0.258) [-0.132, 0.878] & -0.273 (0.236) [-0.75, 0.204] & 0.844 (0.254) [0.404, 1.324] & 0.464 (0.225) [0.16, 0.857] & 0.324 (0.26) [-0.169, 0.839] & -0.287 (0.24) [-0.765, 0.149] \\ 
\hline
\makecell{12. if \\missing\\ on \\ favorable \\of China} & 0.832 (0.245) [0.458, 1.508] & 0.316 (0.135) [0.134, 0.746] & 0.484 (0.259) [-0.024, 0.991] & -0.25 (0.236) [-0.726, 0.226] & 0.936 (0.303) [0.434, 1.552] & 0.452 (0.243) [0.118, 0.933] & 0.398 (0.28) [-0.134, 0.949] & -0.244 (0.257) [-0.73, 0.229] \\
\hline
\hline
\insertTableNotes\\
\label{tab3:ArabBarometerFullResults_addCov}
\end{longtable}
\endgroup
\end{ThreePartTable}
\end{landscape}

\subsection*{Appendix E: Outcome Variables Used in the Health and Retirement Study}
\label{Append:HRSOutcomes}
\begin{ThreePartTable}
\begingroup\fontsize{12pt}{12pt}\selectfont
\begin{longtable}{l  c  l }
\caption{Outcome Variables Used in the Health and Retirement Study} \\ 
  \hline
\hline
Questions & Original response categories & \makecell{Response categories \\used in the study }\\ 
  \hline
\hline
\multicolumn{3}{c}{CESD questions}\\
\hline
\makecell{Much of the time \\during the past week, \\you felt depressed.} & \makecell{1. Yes\\ 5. No} 
 &  \makecell{1. Yes\\ 0. No}\\ 
\hline
\makecell{Much of the time \\during the past week, \\you felt that everything \\you did was an
         effort.} & \makecell{1.Yes\\ 5. No}  & \makecell{1. Yes\\ 0. No}  \\ 
\hline
\makecell{Much of the time \\during the past week, \\your sleep was restless.} & \makecell{1.Yes\\ 5. No}  & \makecell{1. Yes\\ 0. No}  \\ 
\hline
\makecell{Much of the time \\during the past week, \\you were happy.} & \makecell{1.Yes\\ 5. No}  & \makecell{1. Yes\\ 0. No}  \\ 
\hline
\makecell{Much of the time \\during the past week, \\you felt lonely.} & \makecell{1.Yes\\ 5. No}  & \makecell{1. Yes\\ 0. No}  \\  
\hline
\makecell{Much of the time \\during the past week, \\you enjoyed life.} & \makecell{1.Yes\\ 5. No}  & \makecell{1. Yes\\ 0. No}  \\  
\hline
\makecell{Much of the time \\during the past week, \\you felt sad.} & \makecell{1.Yes\\ 5. No}  & \makecell{1. Yes\\ 0. No}  \\  
\hline
\makecell{Much of the time \\during the past week, \\you could not get going.} & \makecell{1.Yes\\ 5. No}  & \makecell{1. Yes\\ 0. No}  \\  
\hline
\hline
\multicolumn{3}{c}{Interviewer Observations}\\
\hline
\makecell{How attentive was the \\respondent to the questions \\during the
         interview?} & \makecell{1. Not at all attentive\\
2. Somewhat attentive\\
3. Very attentive
}  & \makecell{1. Very attentive\\
0. Not at all or\\
\quad somewhat attentive
}  \\ 
\hline
\makecell{How was the respondent's\\ understanding of the questions?} & \makecell{1. Excellent\\
2. Good\\
3. Fair\\
4. Poor
}  & \makecell{1. Excellent\\
0. Good, fair, or poor\\
}  \\ 
\hline
\makecell{How was the respondent's\\ cooperation during the interview?} & \makecell{1. Excellent\\
2. Good\\
3. Fair\\
4. Poor
}  & \makecell{1. Excellent\\
0. Good, fair, or poor\\
}  \\
\hline
\makecell{How much difficulty did \\the respondent have \\remembering things that \\you asked him/her about?} & \makecell{1. No difficulty\\
2. A little difficulty\\
3. Some difficulty\\
4. A lot of difficulty\\
5. Could not do at all
}  & \makecell{1. No difficulty\\
0. A little/some/lot of difficulty\\or could not do at all\\
}  \\
\hline
\makecell{How much difficulty did \\the respondent have hearing\\you when you talked \\to him/her?} & \makecell{1. No difficulty\\
2. A little difficulty\\
3. Some difficulty\\
4. A lot of difficulty\\
5. Could not do at all
}  & \makecell{1. No difficulty\\
0. A little/some/lot of difficulty\\or could not do at all
}   \\
\hline
\makecell{Overall, what is your \\opinion of the quality \\of this interview?\\Was it of: } & \makecell{1. High quality\\
2. Adequate quality\\
3. Questionable quality
}  & \makecell{1. High quality\\
0. Adequate \\or questionable quality
}   \\
\hline
\hline
\multicolumn{3}{c}{Physical activity}\\
\hline
\makecell{How often do you \\take part in sports \\or activities that are\\
         vigorous, such as running \\or jogging, swimming, \\cycling, aerobics or gym\\
         workout, tennis, or \\digging with a spade or shovel} & \makecell{1. More than once a week\\
2. Once a week\\
3. One to three times a month\\
4. Hardly ever or never\\
7. (VOL) Every day 
}  & \makecell{1. At least once a week\\
0. Less than once a week
}   \\
\hline
\makecell{And how often do you \\take part in sports \\or activities that are \\moderately
         energetic such as, \\gardening, cleaning the car, \\walking at a moderate pace,\\
         dancing, floor or \\stretching exercises:} & \makecell{1. More than once a week\\
2. Once a week\\
3. One to three times a month\\
4. Hardly ever or never\\
7. (VOL) Every day 
}  & \makecell{1. At least once a week\\
0. Less than once a week
}\\ 
\hline
\makecell{And how often do \\you take part in \\sports or activities that \\are mildly
         energetic, \\such as vacuuming, \\laundry, home repairs:} & \makecell{1. More than once a week\\
2. Once a week\\
3. One to three times a month\\
4. Hardly ever or never\\
7. (VOL) Every day 
}  & \makecell{1. At least once a week\\
0. Less than once a week}\\
\hline
\hline
\label{tab3:HRS_variable_info}
\end{longtable}
\endgroup
\end{ThreePartTable}

\subsection*{Appendix F: Full Results on Interviewer Variances in the Health and Retirement Study}
\begin{landscape}
\begin{ThreePartTable}
\begin{TableNotes}
\small
    \item Notes: $\beta_1$ is the mode effects in means, computed as the mean of the FTF estimate minus the mean of the TEL estimate. $\sigma^2_f$ is the FTF interviewer variances. $\sigma^2_t$ is the TEL interviewer variance. $rho_{f,int}$ is the interviewer intraclass correlation associated with the FTF mode. $rho_{t,int}$ is the interviewer intraclass correlation associated with the TEL mode. $\alpha$ refers to the log differences between the FTF and TEL interviewer variances. $\rho$ is the correlation between the FTF and TEL random interviewer effects. We use \say{N/A} to mask estimates that cannot be estimated due to numerical difficulties. 
\end{TableNotes}
\begingroup\fontsize{11pt}{11pt}\selectfont
\begin{longtable}{
@{}
>{\raggedright}p{0.09\linewidth}
>{\raggedright}p{0.07\linewidth}
>{\raggedright}p{0.07\linewidth}
>{\raggedright}p{0.07\linewidth}
>{\raggedright}p{0.07\linewidth}
>{\raggedright}p{0.07\linewidth}
>{\raggedright}p{0.07\linewidth}
>{\raggedright}p{0.07\linewidth}
>{\raggedright}p{0.07\linewidth}
>{\raggedright}p{0.07\linewidth}
p{0.07\linewidth}
@{}}
\caption{Interviewer Variances Per Mode for Selected Items in Health and Retirement Study Adjusting for Covariates} \\ 
  \hline
\hline
\multirow{2}{*}{Questions} & \multicolumn{4}{c}{Likelihood} & \multicolumn{4}{c}{Bayesian} \\ 
  \cmidrule(lr){2-6}
  \cmidrule(lr){7-11}
& $\sigma^2_f$ & $\sigma^2_t$ & $\alpha$ & $\beta_1$ &$\rho$ & $\sigma^2_f$ & $\sigma^2_t$ & $\alpha$ & $\beta_1$ &$\rho$ \\
\hline
\hline
\multicolumn{11}{c}{CESD questions}\\
\hline
\makecell{felt \\depressed} & 0.011 (0.006) [0.004, 0.031] & 0.014 (0.008) [0.004, 0.045] & -0.095 (0.389) [-0.857, 0.667] & 0.056 (0.029) [-0.001, 0.113] & 0.222 (0.448) [-0.603, 0.818] & 0.011 (0.006) [0, 0.022] & 0.013 (0.009) [0, 0.03] & 0.044 (0.643) [-1.148, 1.533] & \textbf{0.056} (0.029) [0.005, 0.114] & 0.07 (0.391) [-0.551, 0.874]  \\ 
\hline
\makecell{everything\\ was an\\
         effort} & 0.022 (0.006) [0.013, 0.037] & 0.004 (0.005) [0, 0.052] & 0.893 (0.682) [-0.445, 2.23] & \textbf{0.116} (0.025) [0.066, 0.165] & -0.099 (0.618) [-0.867, 0.809] & 0.025 (0.014) [0.013, 0.037] & 0.007 (0.005) [0.001, 0.016] & 0.746 (0.38) [-0.002, 1.496] & \textbf{0.118} (0.029) [0.071, 0.175] & -0.128 (0.264) [-0.56, 0.254] \\ 
\hline
\makecell{restless\\ sleep} & 0.003 (0.003) [0, 0.02] & 0.004 (0.004) [0, 0.032] & -0.13 (0.719) [-1.54, 1.279] & \textbf{0.057} (0.022) [0.013, 0.1] & -0.698 (1.018) [-1, 0.995] & 0.002 (0.002) [0, 0.007] & 0.005 (0.004) [0, 0.012] & -0.486 (0.754) [-1.89, 0.925] & \textbf{0.053} (0.021) [0.011, 0.095] & 0.337 (0.312) [-0.162, 0.849] \\ 
\hline
\makecell{happy} & 0.007 (0.015) [0, 0.019] & 0.010 (0.014) [0, 0.023] & -0.253 (1.519) [-2.348, 2.915] & 0.033 (0.032) [-0.019, 0.085] & NA (NA) [NA, NA] & 0.011 (0.005) [0.003, 0.021] & 0.011 (0.007) [0, 0.022] & 0.128 (0.539) [-0.889, 1.333] & 0.032 (0.027) [-0.024, 0.083] & -0.518 (0.314) [-0.989, -0.006] \\
\hline
\makecell{lonely}  & 0.006 (0.004) [0.001, 0.025] & 0.004 (0.006) [0, 0.12] & 0.223 (0.973) [-1.685, 2.131] & 0.046 (0.026) [-0.004, 0.097] & -0.208 (1.053) [-0.983, 0.96] & 0.006 (0.004) [0, 0.014] & 0.006 (0.005) [0, 0.016] & 0.178 (0.878) [-1.455, 1.846] & 0.048 (0.028) [-0.005, 0.099] & 0.055 (0.084) [-0.108, 0.218] \\
\hline
\makecell{enjoyed\\ life} & 0.009 (0.006) [0.002, 0.037] & 0.011 (0.009) [0.002, 0.052] & -0.124 (0.528) [-1.16, 0.911] & \textbf{0.07} (0.033) [0.006, 0.134] & -0.823 (0.718) [-1, 0.997] & 0.01 (0.006) [0.001, 0.021] & 0.007 (0.009) [0, 0.025] & 0.551 (0.944) [-1.096, 2.148] & 0.061 (0.036) [-0.005, 0.134] & \textbf{0.56} (0.187) [0.223, 0.921] \\ 
\hline
\makecell{felt\\ sad} &0.03 (0.007) [0.018, 0.048] & 0 (0.001) [0, 1.263] & 2.475 (2.213) [-1.863, 6.813] & 0.047 (0.025) [-0.003, 0.097] & NA (NA) [NA, NA] & 0.032 (0.008) [0.018, 0.048] & 0.003 (0.003) [0, 0.009] & \textbf{1.694} (0.951) [0.463, 3.775] & 0.046 (0.027) [-0.01, 0.097] & \textbf{0.296} (0.137) [0.037, 0.577] \\
\hline
\makecell{could \\not get\\ going} & 0.016 (0.005) [0.008, 0.03] & 0.019 (0.008) [0.008, 0.044] & -0.098 (0.274) [-0.635, 0.439] & 0.05 (0.027) [-0.003, 0.103] & 0.314 (0.318) [-0.351, 0.768] & 0.02 (0.032) [0.007, 0.029] & 0.02 (0.008) [0.006, 0.035] & -0.051 (0.339) [-0.732, 0.515] & 0.051 (0.033) [-0.01, 0.109] & 0.274 (0.307) [-0.346, 0.797] \\
\hline
\makecell{overall\\ indicator} & 0.012 (0.005) [0.005, 0.029] & 0.012 (0.007) [0.004, 0.04] & 0.006 (0.37) [-0.718, 0.731] & \textbf{0.152} (0.028) [0.096, 0.207] & -0.264 (0.428) [-0.825, 0.558] & 0.016 (0.028) [0.002, 0.024] & 0.012 (0.008) [0.001, 0.027] & 0.093 (0.487) [-0.901, 1.075] & \textbf{0.15} (0.027) [0.102, 0.207] & 0.244 (0.211) [-0.18, 0.573] \\
\hline
\multicolumn{11}{c}{Interviewer Observations}\\
\hline
\makecell{attentive}  & 0.29 (0.032) [0.233, 0.361] & 0.342 (0.043) [0.268, 0.438] & -0.082 (0.063) [-0.205, 0.041] & 0.013 (0.035) [-0.056, 0.081] & \textbf{0.893} (0.036) [0.795, 0.946] & 0.298 (0.032) [0.233, 0.356] & 0.351 (0.044) [0.262, 0.431] & -0.081 (0.062) [-0.197, 0.038] & 0.018 (0.035) [-0.049, 0.088] & \textbf{0.878} (0.039) [0.803, 0.955] \\
\hline
\makecell{understanding} & 0.408 (0.04) [0.336, 0.494] & 0.461 (0.05) [0.373, 0.571] & -0.062 (0.049) [-0.158, 0.033] & -0.003 (0.032) [-0.066, 0.059] & \textbf{0.921} (0.023) [0.86, 0.956] & 0.413 (0.039) [0.341, 0.493] & 0.465 (0.051) [0.366, 0.56] & -0.058 (0.049) [-0.149, 0.043] & 0 (0.032) [-0.064, 0.061] & \textbf{0.91} (0.041) [0.861, 0.958] \\
\hline
\makecell{cooperation} & 0.45 (0.043) [0.373, 0.542] & 0.404 (0.044) [0.327, 0.5] & 0.053 (0.047) [-0.039, 0.145] & 0.174 (0.031) [0.113, 0.234] & \textbf{0.941} (0.021) [0.884, 0.971] & 0.459 (0.047) [0.378, 0.556] & 0.41 (0.048) [0.321, 0.51] & 0.057 (0.047) [-0.039, 0.138] & 0.178 (0.032) [0.108, 0.236] & \textbf{0.931} (0.025) [0.881, 0.971] \\
\hline
\makecell{remembering}&0.482 (0.047) [0.398, 0.584] & 0.593 (0.065) [0.478, 0.735] & \textbf{-0.103} (0.047) [-0.195, -0.011] & \textbf{-0.065} (0.032) [-0.128, -0.001] & \textbf{0.941} (0.019) [0.89, 0.969] & 0.483 (0.047) [0.392, 0.574] & 0.605 (0.059) [0.489, 0.721] & \textbf{-0.112} (0.047) [-0.205, -0.028] & \textbf{-0.062} (0.033) [-0.124, 0.002] & \textbf{0.931} (0.029) [0.885, 0.972] \\
\hline
\makecell{hearing} & 0.27 (0.03) [0.217, 0.336] & 0.372 (0.046) [0.291, 0.476] & \textbf{-0.161} (0.063) [-0.285, -0.038] & \textbf{0.152} (0.034) [0.084, 0.219] & \textbf{0.888} (0.035) [0.796, 0.94] & 0.271 (0.032) [0.212, 0.335] & 0.375 (0.048) [0.274, 0.462] & \textbf{-0.161} (0.065) [-0.284, -0.037] & \textbf{0.151} (0.038) [0.084, 0.229] & \textbf{0.87} (0.064) [0.795, 0.947] \\
\hline
\makecell{Overall\\ quality} & 0.879 (0.08) [0.736, 1.051] & 0.782 (0.079) [0.642, 0.953] & 0.058 (0.04) [-0.019, 0.136] & \textbf{0.09} (0.034) [0.023, 0.156] & \textbf{0.96} (0.014) [0.923, 0.98] & 0.881 (0.077) [0.749, 1.04] & 0.788 (0.08) [0.641, 0.949] & 0.057 (0.046) [-0.032, 0.14] & \textbf{0.086} (0.038) [0.014, 0.158] & \textbf{0.94} (0.088) [0.913, 0.983] \\
\hline
\hline
\multicolumn{11}{c}{Physical activity}\\
\hline
\makecell{vigorous\\ sports} & 0.015 (0.004) [0.008, 0.027] & 0.008 (0.006) [0.002, 0.032] & 0.298 (0.381) [-0.45, 1.045] & -0.036 (0.022) [-0.079, 0.007] & 0.642 (0.41) [-0.541, 0.972] & 0.017 (0.011) [0.007, 0.026] & 0.007 (0.004) [0, 0.015] & 0.523 (0.406) [-0.209, 1.45] & -0.037 (0.026) [-0.081, 0.014] & 0.36 (0.372) [-0.446, 0.827] \\
\hline
\makecell{moderately\\ energetic\\ sports} & 0.013 (0.004) [0.007, 0.026] & 0.015 (0.006) [0.006, 0.035] & -0.043 (0.274) [-0.581, 0.494] & 0.028 (0.023) [-0.017, 0.073] & 0.478 (0.326) [-0.297, 0.873] & 0.015 (0.008) [0.006, 0.024] & 0.019 (0.008) [0.004, 0.033] & -0.086 (0.28) [-0.655, 0.464] & 0.031 (0.025) [-0.019, 0.078] & 0.233 (0.248) [-0.351, 0.698] \\
\hline
\makecell{mildly\\ energetic\\ sports}  & 0.015 (0.005) [0.007, 0.03] & 0.03 (0.009) [0.016, 0.056] & -0.353 (0.238) [-0.819, 0.114] & \textbf{0.135} (0.028) [0.08, 0.19] & 0.107 (0.278) [-0.416, 0.577] & 0.02 (0.042) [0.002, 0.03] & 0.031 (0.01) [0.014, 0.052] & -0.355 (0.361) [-1.097, 0.324] & \textbf{0.134} (0.028) [0.073, 0.184] & 0.144 (0.29) [-0.264, 0.962] \\ 
\hline
\hline
 \insertTableNotes\\
\label{tab3:HRSFullResults_addCov}
\end{longtable}
\endgroup
\end{ThreePartTable}
\end{landscape}

\subsection*{Appendix G: Results Testing Sensitivity to Rho for the Depression Items in the Health and Retirement Study}
\begin{landscape}
\begin{ThreePartTable}
\begin{TableNotes}
\small
    \item Notes: $\beta_1$ is the mode effects in means, computed as the mean of the FTF estimate minus the mean of the TEL estimate. $\sigma^2_f$ is the FTF interviewer variances. $\sigma^2_t$ is the TEL interviewer variance. $\alpha$ refers to the log differences between the FTF and TEL interviewer variances. $\rho$ is the correlation between the FTF and TEL random interviewer effects. We use \say{N/A} to mask estimates that are unstable (with a standard error greater than 5) or cannot be estimated due to numerical difficulties. 
\end{TableNotes}
\begingroup\fontsize{11pt}{11pt}\selectfont
\begin{longtable}{
@{}
>{\raggedright}p{0.09\linewidth}
>{\raggedright}p{0.07\linewidth}
>{\raggedright}p{0.07\linewidth}
>{\raggedright}p{0.07\linewidth}
>{\raggedright}p{0.07\linewidth}
>{\raggedright}p{0.07\linewidth}
>{\raggedright}p{0.07\linewidth}
>{\raggedright}p{0.07\linewidth}
p{0.07\linewidth}
@{}}
\caption{Interviewer Variances Per Mode for Depression Items in Health and Retirement Study Adjusting for Covariates Using Likelihood Estimation} \\ 
  \hline
\hline
\multirow{2}{*}{Questions} & \multicolumn{4}{c}{$\rho=0$} & \multicolumn{4}{c}{$\rho=\hat{\rho}_{Bayesian}$} \\ 
  \cmidrule(lr){2-5}
  \cmidrule(lr){6-9}
& $\sigma^2_f$ & $\sigma^2_t$ & $\alpha$ & $\beta_1$ & $\sigma^2_f$ & $\sigma^2_t$ & $\alpha$ & $\beta_1$\\
\hline
\hline
\makecell{felt \\depressed} & 0.011 (0.006) [0.004, 0.031] & 0.014 (0.008) [0.004, 0.045] & -0.094 (0.391) [-0.861, 0.674] & 0.056 (0.029) [-0.002, 0.114] & 0.011 (0.006) [0.004, 0.031] & 0.014 (0.008) [0.004, 0.045] & -0.095 (0.387) [-0.854, 0.663] & 0.056 (0.029) [-0.001, 0.114] \\ 
\hline
\makecell{everything\\ was an\\
         effort} & 0.022 (0.006) [0.013, 0.037] & 0.004 (0.005) [0, 0.053] & 0.893 (0.686) [-0.452, 2.238] & \textbf{0.115} (0.025) [0.066, 0.164] & 0.022 (0.006) [0.013, 0.037] & 0.004 (0.005) [0, 0.052] & 0.897 (0.685) [-0.445, 2.238] & \textbf{0.116} (0.025) [0.066, 0.165] \\  
\hline
\makecell{restless\\ sleep} & 0.003 (0.003) [0, 0.021] & 0.004 (0.004) [0, 0.032] & -0.157 (0.735) [-1.598, 1.284] & \textbf{0.057} (0.022) [0.014, 0.099] & 0.002 (0.003) [0, 0.027] & 0.003 (0.004) [0, 0.042] & -0.159 (0.88) [-1.883, 1.565] & \textbf{0.057} (0.021) [0.015, 0.099] \\ 
\hline
\makecell{happy} & 0.01 (0.006) [0.003, 0.032] & 0.008 (0.007) [0.001, 0.048] & 0.124 (0.559) [-0.971, 1.22] & 0.03 (0.027) [-0.023, 0.083] & 0.011 (0.006) [0.004, 0.03] & 0.01 (0.007) [0.003, 0.037] & 0.029 (0.416) [-0.786, 0.844] & 0.032 (0.028) [-0.023, 0.087] \\
\hline
\makecell{lonely}  & 0.005 (0.004) [0.001, 0.025] & 0.004 (0.006) [0, 0.117] & 0.216 (0.971) [-1.687, 2.12] & 0.046 (0.026) [-0.004, 0.096] & 0.005 (0.004) [0.001, 0.025] & 0.003 (0.006) [0, 0.124] & 0.222 (0.987) [-1.713, 2.158] & 0.046 (0.025) [-0.004, 0.096] \\
\hline
\makecell{enjoyed\\ life} & 0.009 (0.006) [0.002, 0.037] & 0.012 (0.009) [0.003, 0.052] & -0.122 (0.521) [-1.143, 0.9] & \textbf{0.069} (0.032) [0.007, 0.131] & 0.006 (0.007) [0.001, 0.062] & 0.007 (0.01) [0, 0.123] & -0.034 (0.929) [-1.855, 1.787] & \textbf{0.064} (0.031) [0.003, 0.125] \\
\hline
\makecell{felt\\ sad} & 0.03 (0.007) [0.018, 0.047] & N/A (N/A) [N/A, N/A] & N/A (N/A) [N/A, N/A] & 0.048 (0.026) [-0.003, 0.099] & 0.03 (0.007) [0.018, 0.047] & N/A (N/A) [N/A, N/A] & 2.123 (2.982) [-3.721, 7.967] & 0.048 (0.026) [-0.003, 0.099] \\
\hline
\makecell{could \\not get\\ going} & 0.016 (0.005) [0.008, 0.03] & 0.019 (0.008) [0.008, 0.044] & -0.09 (0.275) [-0.628, 0.449] & 0.051 (0.028) [-0.003, 0.106] & 0.016 (0.005) [0.008, 0.03] & 0.019 (0.008) [0.008, 0.044] & -0.099 (0.273) [-0.633, 0.436] & 0.05 (0.027) [-0.003, 0.103] \\
\hline
\makecell{overall\\ indicator} & 0.012 (0.005) [0.005, 0.029] & 0.012 (0.007) [0.004, 0.04] & 0.009 (0.374) [-0.723, 0.741] & \textbf{0.152} (0.028) [0.097, 0.206] & 0.012 (0.005) [0.005, 0.028] & 0.011 (0.007) [0.003, 0.041] & 0.038 (0.398) [-0.742, 0.819] & \textbf{0.151} (0.027) [0.098, 0.204] \\
\hline
\hline
 \insertTableNotes\\
\label{tab3:HRSResults_addCov_rho0}
\end{longtable}
\endgroup
\end{ThreePartTable}
\end{landscape}

\newpage
\singlespacing

   \bibliographystyle{references/ajs}
   \bibliography{references/main} 















\end{document}